%% file: main.tex
\documentclass{llncs}

\usepackage{makeidx}

\usepackage[T1]{fontenc}

\usepackage{bm}
\usepackage{pgfplots}
\usepackage{enumerate,paralist}
\usepackage{pbox}

\usepackage{parskip}
\usepackage{multirow}
\usepackage{url}
\usepackage{graphicx}
\usepackage{wrapfig}
\usepackage{tikz,pgffor}
\usetikzlibrary{arrows}
\usetikzlibrary{shapes}
\usetikzlibrary{calc}
\usetikzlibrary{automata}
\usetikzlibrary{positioning}

\tikzstyle{label}=[shape=circle,draw,inner sep=0pt,minimum size=5mm]
\tikzstyle{tran}=[draw,->,>=stealth, rounded corners]

\usepackage{lmodern}
\usepackage{amsmath}
\usepackage{amssymb}
\usepackage{listings}
\usepackage{mathtools}
\usepackage{times}
\usepackage{stmaryrd}

\lstdefinelanguage{prog}
{
morekeywords={if, then, else, end, for, true, false, and, or, skip, return},
sensitive = false
}

\DeclareMathAlphabet{\mathpzc}{OT1}{pzc}{m}{it}

\newcommand{\Rset}{\mathbb{R}}
\newcommand{\Nset}{\mathbb{N}}
\newcommand{\Zset}{\mathbb{Z}}

\newcommand{\Uni}[1]{\mathsf{Uni}(#1)}

\begin{document}

\input{introduction}

\input{prelim}

\input{average_case}
\input{synalgorithm}

\input{experiments}

\input{conclu}


\clearpage
\bibliographystyle{splncs03}
\bibliography{PL}

\clearpage
\input{appendix}

\end{document}

%% file: introduction.tex
\title{Automated Recurrence Analysis \\for Almost-Linear Expected-Runtime Bounds}


\author{
Krishnendu Chatterjee\inst{1} \and Hongfei Fu\inst{2} \and Aniket Murhekar\inst{3}
}

\authorrunning{Chatterjee et al.}

\institute{
IST Austria, Klosterneuburg, Austria
\and
State Key Laboratory of Computer Science, Institute of Software\\ Chinese
Academy of Sciences, Beijing, P.R. China
\and
IIT Bombay, India
}

\maketitle

\begin{abstract}
We consider the problem of developing automated techniques for solving recurrence relations to
aid the expected-runtime analysis of programs.
Several classical textbook algorithms have quite efficient expected-runtime
complexity, whereas the corresponding worst-case bounds are
either inefficient (e.g., {\sc Quick-Sort}), or
completely ineffective (e.g., {\sc Coupon-Collector}).
Since the main focus of expected-runtime analysis is to obtain
efficient bounds, we consider bounds that are either
logarithmic, linear, or almost-linear
($\mathcal{O}(\log{n})$, $\mathcal{O}(n)$, $\mathcal{O}(n\cdot\log{n})$,
respectively, where $n$ represents the input size).
Our main contribution is an efficient (simple linear-time algorithm)
sound approach for deriving such expected-runtime bounds for the analysis of
recurrence relations induced by randomized algorithms.
Our approach can infer the asymptotically optimal expected-runtime bounds
for recurrences of classical randomized algorithms, including
{\sc Randomized-Search, Quick-Sort, Quick-Select, Coupon-Collector},
where the worst-case bounds are either inefficient
(such as linear as compared to logarithmic of expected-runtime, or quadratic as
compared to linear or almost-linear of expected-runtime), or ineffective.
We have implemented our approach, and the experimental results show
that we obtain the bounds efficiently for the recurrences of
various classical algorithms.
\end{abstract}

\vspace{-1.5em}
\section{Introduction}
\vspace{-1em}

\noindent{\em Static analysis for quantitative bounds.}
Static analysis of programs aims to reason about programs without
running them.
The most basic properties for static analysis are qualitative properties,
such as safety, termination, liveness, that for every trace of a program
gives a Yes or No answer (such as assertion violation or not, termination
or not).
However, recent interest in analysis of resource-constrained systems,
such as embedded systems, as well as for performance analysis,
quantitative performance characteristics are necessary.
For example, the qualitative problem of termination asks whether a given
program always terminates, whereas the quantitative problem asks to
obtain precise bounds on the number of steps, and is thus a more
challenging problem.
Hence the problem of automatically reasoning about resource bounds (such
as time complexity bounds) of programs is both of significant
theoretical as well as practical interest.

\smallskip\noindent{\em Worst-case bounds.}
The worst-case analysis of programs is the fundamental problem in computer
science, which is the basis of algorithms and complexity theory.
However, manual proofs of worst-case analysis can be tedious and also require
non-trivial mathematical ingenuity, e.g., the book {\em The Art of Computer Programming}
by Knuth presents a wide range of involved techniques to derive such precise
bounds~\cite{DBLP:books/aw/Knuth73a,DBLP:books/aw/Knuth73}.
There has been a considerable research effort for automated analysis of such
worst-case bounds for programs, see~\cite{SPEED1,SPEED2,Hoffman1,Hoffman2}
for excellent expositions on the significance of deriving precise worst-case
bounds and the automated methods to derive them.
For the worst-case analysis there are several techniques, such as
worst-case execution time analysis~\cite{DBLP:journals/tecs/WilhelmEEHTWBFHMMPPSS08},  resource analysis using abstract
interpretation and type systems~\cite{SPEED2,DBLP:journals/entcs/AlbertAGGPRRZ09,DBLP:conf/popl/JostHLH10,Hoffman1,Hoffman2},  ranking functions~\cite{BG05,DBLP:conf/cav/BradleyMS05,DBLP:conf/tacas/ColonS01,DBLP:conf/vmcai/PodelskiR04,DBLP:conf/pods/SohnG91,DBLP:conf/vmcai/Cousot05,DBLP:journals/fcsc/YangZZX10,DBLP:journals/jossac/ShenWYZ13},
as well as recurrence relations~\cite{DBLP:conf/icfp/Grobauer01,DBLP:journals/entcs/AlbertAGGPRRZ09,DBLP:conf/sas/AlbertAGP08,DBLP:conf/esop/AlbertAGPZ07}.

\smallskip\noindent{\em Expected-runtime bounds.}
While several works have focused on deriving worst-case bounds for programs,
quite surprisingly little work has been done to derive precise bounds for
expected-runtime analysis, with the exception of~\cite{DBLP:journals/tcs/FlajoletSZ91},
which focuses on randomization in combinatorial structures (such as trees).
This is despite the fact that expected-runtime analysis is an equally important
pillar of theoretical computer science, both in terms of theoretical and
practical significance.
For example, while for real-time systems with hard constraints worst-case
analysis is necessary, for real-time systems with soft constraints the more
relevant information is the expected-runtime analysis.
Below we highlight three key significance of expected-runtime analysis.
\begin{compactenum}
\item {\em Simplicity and desired properties:}
The first key aspect is {\em simplicity}: often much simpler
algorithms (thus simple and efficient implementations) exist
for expected-runtime complexity as compared to worst-case complexity.
A classic example is the {\sc Selection} problem that given a set of $n$
numbers and $0\leq k \leq n$, asks to find the $k$-th largest number
(eg, for median $k=n/2$).
The classical linear-time algorithm for the problem (see~\cite[Chapter~9]{DBLP:books/daglib/0023376})
is quite involved, and its worst-case analysis to obtain linear time bound is
rather complex.
In contrast, a much simpler algorithm exists (namely, {\sc Quick-Select})
that has linear expected-runtime complexity.
Moreover, randomized algorithms with expected-runtime complexity enjoy
many desired properties, which deterministic algorithms do not have.
A basic example is {\sc Channel-Conflict Resolution}
(see Example~\ref{ex:channel}, Section~\ref{sec:motivatingbi})
where the simple randomized algorithm can be implemented in a distributed or
concurrent setting, whereas deterministic algorithms are quite cumbersome.

\item {\em Efficiency in practice:}
Since worst-case analysis concerns with corner cases that rarely arise,
many algorithms and implementations have much better expected-runtime complexity,
and they perform extremely well in practice.
A classic example is the {\sc Quick-Sort} algorithm, that has quadratic worst-case
complexity, but almost linear expected-runtime complexity, and is one of the most
efficient sorting algorithms in practice.

\item {\em Worst-case analysis ineffective:}
In several important cases
the worst-case analysis is completely ineffective.
For example, consider one of the textbook stochastic process, namely the
{\sc Coupon-Collector} problem, where there are $n$ types of coupons to be collected,
and in each round, a coupon type among the $n$ types is obtained uniformly at random.
The process stops when all types are collected.
The {\sc Coupon-Collector} process is one of the basic and classical stochastic
processes, with numerous applications in network routing, load balancing,
etc (see~\cite[Chapter 3]{DBLP:books/cu/MotwaniR95} for applications of
{\sc Coupon-Collector} problems).
For the worst-case analysis, the process might not terminate (worst-case bound infinite),
but the expected-runtime analysis shows that the expected termination
time is $\mathcal{O}(n \cdot \log n)$.
\end{compactenum}

\smallskip\noindent{\em Challenges.}
The expected-runtime analysis brings several new challenges as compared to
the worst-case analysis.
First, for the worst-case complexity bounds, the most classical characterization
for analysis of recurrences is the {\em Master Theorem} (cf.~\cite[Chapter~1]{DBLP:books/daglib/0023376}) and Akra-Bazzi's Theorem~\cite{DBLP:journals/coap/AkraB98}.
However, the expected-runtime analysis problems give rise to recurrences that
are not characterized by these theorems since our recurrences normally involve an unbounded summation resulting from a randomized selection of integers from $1$ to $n$ where $n$ is unbounded.
Second, techniques like ranking functions (linear or polynomial ranking
functions) cannot derive efficient bounds such as $\mathcal{O}(\log n)$
or $\mathcal{O}(n \cdot \log n)$.
While expected-runtime analysis has been considered for combinatorial structures
using generating function~\cite{DBLP:journals/tcs/FlajoletSZ91},
we are not aware of any automated technique to handle recurrences arising from
randomized algorithms.

\smallskip\noindent{\em Analysis problem.}
We consider the algorithmic analysis problem of recurrences
arising naturally for randomized recursive programs.
Specifically we consider the following:
\begin{compactitem}
\item We consider two classes of recurrences:
(a)~{\em univariate} class with one variable (which represents the array
length, or the number of input elements, as required in problems such as
{\sc Quick-Select, Quick-Sort} etc); and
(b)~{\em separable bivariate} class with two variables (where the two independent
variables represent the total number of elements and total number of successful cases,
respectively, as required in problems such as {\sc Coupon-Collector, Channel-Conflict Resolution}).
The above two classes capture a large class of expected-runtime analysis problems,
including all the classical ones mentioned above.
Moreover, the main purpose of expected-runtime analysis is to obtain efficient bounds.
Hence we focus on the case of logarithmic, linear, and almost-linear bounds
(i.e., bounds of form $\mathcal{O}(\log n)$, $\mathcal{O}(n)$ and
$\mathcal{O}(n \cdot \log n)$, respectively, where $n$ is the size of the
input).
Moreover, for randomized algorithms, quadratic bounds or higher are rare.
\end{compactitem}
Thus the main problem we consider is to automatically derive such efficient bounds
for randomized univariate and separable bivariate recurrence relations.

\smallskip\noindent{\em Our contributions.}
Our main contribution is a sound approach for analysis of recurrences for expected-runtime analysis.
The input to our problem is a recurrence relation and the output is either
logarithmic, linear, or almost-linear as the asymptotic bound, or fail.
The details of our contributions are as follows:
\begin{compactenum}
\item {\em Efficient algorithm.}
We first present a linear-time algorithm for the univariate case, which is
based on simple comparison of leading terms of pseudo-polynomials.
Second, we present a simple reduction for separable bivariate
recurrence analysis to the univariate case.
Our efficient (linear-time) algorithm can soundly infer logarithmic,
linear, and almost-linear bounds for recurrences of
one or two variables.

\item {\em Analysis of classical algorithms.}
We show that for several classical algorithms, such as
{\sc Randomized-Search, Quick-Select, Quick-Sort, Coupon-Collector, Channel-Conflict Resolution}
(see Section~\ref{sec:motivatinguni} and Section~\ref{sec:motivatingbi}
for examples), our sound approach can obtain the asymptotically optimal
expected-runtime bounds for the recurrences.
In all the cases above, either the worst-case bounds (i)~do not exist
(e.g., {\sc Coupon-Collector}), or
(ii)~are quadratic when the expected-runtime bounds are linear or almost-linear
(e.g., {\sc Quick-Select, Quick-Sort});
or (iii)~are linear when the expected-runtime bounds are logarithmic
(e.g., {\sc Randomized-Search}).
Thus in cases where the worst-case bounds are either not applicable,
or grossly overestimate the expected-runtime bounds, our technique is both
efficient (linear-time) and can infer the optimal bounds.

\item {\em Implementation.}
Finally, we have implemented our approach, and we present experimental results on the
classical examples to show that we can efficiently achieve the automated expected-runtime
analysis of randomized recurrence relations.
\end{compactenum}

\noindent{\em Novelty and technical contribution.}
The key novelty of our approach is an automated method to analyze
recurrences arising from randomized recursive programs, which are
not covered by Master theorem.
Our approach is based on a guess-and-check technique.
We show that by over-approximating terms in a recurrence relation through
integral and Taylor's expansion, we can soundly infer logarithmic, linear and
almost-linear bounds using simple comparison between leading terms of
pseudo-polynomials.

%% file: prelim.tex
\newcommand{\eql}{\mathsf{eq}}
\newcommand{\Sub}{\mathsf{Subst}}
\newcommand{\Eval}{\mathsf{Eval}}

\vspace{-1.5em}
\section{Recurrence Relations}\label{sec:recurrel}
\vspace{-1em}
We present our mini specification language for recurrence
relations for expected-runtime analysis.
The language is designed to capture running time of recursive randomized
algorithms which involve (i)~only one function call whose expected-runtime
complexity is to be determined, (ii)~at most two integer parameters,
and (iii)~involve randomized-selection or divide-and-conquer techniques.
We present our language separately for the univariate and bivariate cases.
In the sequel, we denote by $\Nset$, $\Nset_0$, $\Zset$, and $\Rset$ the
sets of all positive integers, non-negative integers, integers, and real
numbers, respectively.

\vspace{-1.5em}
\subsection{Univariate Randomized Recurrences}\label{sect:univariate}
\vspace{-1em}
Below we define the notion of univariate randomized recurrence relations.
First, we introduce the notion of univariate recurrence expressions.
Since we only consider single recursive function call, we use `$\mathrm{T}$'
to represent the (only) function call.
We also use `$\mathfrak{n}$' to represent the only parameter in the function
declaration.

\smallskip\noindent{\bf Univariate recurrence expressions.}
The syntax of \emph{univariate recurrence expressions} $\mathfrak{e}$ is
generated by the following grammar:
\begin{align*}
\mathfrak{e} & ::=  c\mid \mathfrak{n}\mid \ln{\mathfrak{n}} \mid \mathfrak{n}\cdot \ln{\mathfrak{n}}\mid \frac{1}{\mathfrak{n}}\mid \mathrm{T}\left(\mathfrak{n}-1\right) \mid \mathrm{T}\left(\left\lfloor\frac{\mathfrak{n}}{2}\right\rfloor\right) \mid \mathrm{T}\left(\left\lceil\frac{\mathfrak{n}}{2}\right\rceil\right)\\
&\mid \frac{\sum_{\mathfrak{j}=1}^{\mathfrak{n}-1} \mathrm{T}(\mathfrak{j})}{\mathfrak{n}}\mid \frac{1}{\mathfrak{n}}\cdot\left( \textstyle\sum_{\mathfrak{j}=\left\lceil\mathfrak{n}/2\right\rceil}^{\mathfrak{n}-1}\mathrm{T}(\mathfrak{j})+ \textstyle\sum_{\mathfrak{j}=\left\lfloor\mathfrak{n}/{2}\right\rfloor}^{\mathfrak{n}-1} \mathrm{T}(\mathfrak{j})\right)\mid c\cdot \mathfrak{e}\mid \mathfrak{e}+\mathfrak{e}
\end{align*}
where $c\in [1,\infty)$ and $\ln(\centerdot)$ represents the natural logarithm function with base $e$.
Informally, $\mathrm{T}(\mathfrak{n})$ is the (expected) running time of a recursive randomized program which involves only one recursive routine indicated by $\mathrm{T}$ and only one parameter indicated by $\mathfrak{n}$.
Then each $\mathrm{T}(\centerdot)$-term in the grammar has a direct algorithmic meaning:
\begin{compactitem}
\item $\mathrm{T}\left(\mathfrak{n}-1\right)$ may mean a recursion to a sub-array with length decremented by one;
\item
$\mathrm{T}\left(\left\lfloor\frac{\mathfrak{n}}{2}\right\rfloor\right)$ and  $\mathrm{T}\left(\left\lceil\frac{\mathfrak{n}}{2}\right\rceil\right)$
may mean a recursion related to a divide-and-conquer technique;
\item finally,
$\frac{\sum_{\mathfrak{j}=1}^{\mathfrak{n}-1} \mathrm{T}(\mathfrak{j})}{\mathfrak{n}}\mbox{ and }\frac{1}{\mathfrak{n}}\cdot\left( \sum_{\mathfrak{j}=\left\lceil\frac{n}{2}\right\rceil}^{\mathfrak{n}-1}\mathrm{T}(\mathfrak{j})+ \sum_{\mathfrak{j}=\left\lfloor\frac{\mathfrak{n}}{2}\right\rfloor}^{\mathfrak{n}-1} \mathrm{T}(\mathfrak{j})\right)$
may mean a recursion related to a randomized selection of an array index.
\end{compactitem}

\smallskip\noindent{\em Substitution.}
Consider a function $h:\Nset\rightarrow\Rset$ and univariate recurrence expression ${\mathfrak{e}}$.
The {\em substitution function}, denoted by $\Sub({\mathfrak{e}},h)$, is the function from $\Nset$ into $\Rset$ such that the value
for $n$ is obtained by evaluation through substituting $h$ for  $\mathrm{T}$ and $n$ for $\mathfrak{n}$ in ${\mathfrak{e}}$, respectively.
Moreover, if $\mathfrak{e}$ does not involve the appearance of `$\mathrm{T}$', then we use the abbreviation
$\Sub({\mathfrak{e}})$ i.e., omit $h$.
For example, (i)~if ${\mathfrak{e}}= \mathfrak{n} + \mathrm{T}(\mathfrak{n}-1)$, and $h: n \mapsto n\cdot \log n$,
then $\Sub({\mathfrak{e}},h)$ is the function $n \mapsto n+ (n-1)\cdot \log (n-1)$,
and (ii)~if ${\mathfrak{e}}= 2\cdot \mathfrak{n}$, then
 $\Sub({\mathfrak{e}})$ is
$n \mapsto 2n$.

\smallskip\noindent{\bf Univariate recurrence relation.}
A {\em univariate recurrence relation} $G=(\eql_1,\eql_2)$ is a pair of equalities as follows:
\begin{equation}\label{eq:unirecurrel}
\eql_1: \ \mathrm{T}(\mathfrak{n})=\mathfrak{e}; \qquad \qquad \eql_2: \ \mathrm{T}(1)=c
\end{equation}
where $c\in (0,\infty)$ and $\mathfrak{e}$ is a univariate recurrence expression.
For a univariate recurrence relation $G$ the {\em evaluation sequence} $\Eval(G)$ is as follows:
$\Eval(G)(1)=c$, and for $n \geq 2$, given $\Eval(G)(i)$ for $1\leq i < n$, for the
value $\Eval(G)(n)$ we evaluate the expression $\Sub(\mathfrak{e},\Eval(G))$,
since in $\mathfrak{e}$ the parameter $\mathfrak{n}$ always decreases and is thus
well-defined.

\smallskip\noindent{\em Finite vs infinite solution.}
Note that the above description gives a computational procedure to compute
$\Eval(G)$ for any finite $n$, in linear time in $n$ through dynamic programming.
The interesting question is to algorithmically analyze the infinite behavior.
A function $T_G:\Nset\rightarrow\Rset$ is called a solution to $G$
if $T_G(n)=\Eval(G)(n)$ for all $n \geq 1$.
The function $T_G$ is unique and explicitly defined as follows:
(1)~\emph{Base Step.} $T_G(1):=c$; and (2)~\emph{Recursive Step.} $T_G(n):=\Sub(\mathfrak{e},T_G)(n)$ for all $n\ge 2$.
The interesting algorithmic question is to reason about the asymptotic infinite behaviour of $T_G$.

\vspace{-1.5em}
\subsection{Motivating Classical Examples}\label{sec:motivatinguni}
\vspace{-1em}
In this section we present several classical examples of randomized
programs whose recurrence relations belong to the
class of univariate recurrence relations described in Section~\ref{sect:univariate}.
We put details of pseudocode and how to derive the recurrence relations in this section in Appendix~\ref{sect:recurreldetails}.
Moreover in all cases the base step is $\mathrm{T}(1)=1$, hence we discuss the recursive case.

\begin{example}[{\sc Randomized-Search}]\label{ex:randsearch}
Consider the Sherwood's {\sc Randomized-Search\ } algorithm (cf.~\cite[Chapter~9]{McConnellbook}).
The algorithm checks whether an integer value $d$ is present within the index range $[i,j]$ ($0\le i\le j$)
in an integer array $ar$ which is sorted in increasing order and is without duplicate entries.
The algorithm outputs either the index for $d$ in $ar$ or $-1$ meaning that $d$ is not present in
the index range $[i,j]$ of  $ar$.
The recurrence relation for this example is as follows:
\begin{equation}\label{eq:relrandsearch}
\textstyle\mathrm{T}(\mathfrak{n})=6+\frac{1}{\mathfrak{n}}\cdot\big( \sum_{\mathfrak{j}=\left\lceil\mathfrak{n}/{2}\right\rceil}^{\mathfrak{n}-1}\mathrm{T}(\mathfrak{j})+ \sum_{\mathfrak{j}=\left\lfloor\mathfrak{n}/{2}\right\rfloor}^{\mathfrak{n}-1} \mathrm{T}(\mathfrak{j})\big)
\end{equation}
We note that the worst-case complexity for this algorithm is $\Theta(n)$.\qed
\end{example}

\begin{example}[{\sc Quick-Sort}]\label{ex:quicksort}
Consider the {\sc Quick-Sort} algorithm~\cite[Chapter~7]{DBLP:books/daglib/0023376}.
The recurrence relation for this example is:
\begin{equation}\label{eq:relquicksort}
\textstyle\mathrm{T}(\mathfrak{n})=2\cdot\mathfrak{n}+ 2\cdot (\sum_{\mathfrak{j}=1}^{\mathfrak{n}-1} \mathrm{T}(\mathfrak{j}))/{\mathfrak{n}}
\end{equation}
where $\mathrm{T}(\mathfrak{n})$ represents the maximal expected execution time where
$\mathfrak{n}$ is the array length and the execution time of {\em pivoting} is represented
by $2\cdot \mathfrak{n}$.
We note that the worst-case complexity for this algorithm is $\Theta(n^2)$.\qed
\end{example}

\begin{example}[{\sc Quick-Select}]\label{ex:quickselect}
Consider the {\sc Quick-Select} algorithm (cf.~\cite[Chapter~9]{DBLP:books/daglib/0023376}).
The recurrence relation for this example is
\begin{equation}\label{eq:relquickselect}
\textstyle\mathrm{T}(\mathfrak{n})\!=\!4+2\cdot\mathfrak{n}+
\frac{1}{\mathfrak{n}}\cdot \left(\sum_{\mathfrak{j}=\left\lfloor \mathfrak{n}/2\right\rfloor}^{\mathfrak{n}-1} \mathrm{T}(\mathfrak{j})+ \sum_{\mathfrak{j}=\left\lceil \mathfrak{n}/2\right\rceil}^{\mathfrak{n}-1} \mathrm{T}(\mathfrak{j})\right)
\end{equation}
We note that the worst-case complexity for this algorithm is $\Theta(n^2)$.\qed
\end{example}

\begin{example}[{\sc Diameter-Computation}]\label{ex:diameter}
Consider the {\sc Diameter-Computation} algorithm (cf.~\cite[Chapter 9]{DBLP:books/cu/MotwaniR95}) to compute the diameter of an input finite set $S$ of three-dimensional points.
Depending on Eucledian or $L_1$ metric we obtain two different recurrence relations.
For Eucledian we have the following relation:
\begin{equation}\label{eq:reldiametera}
\textstyle\mathrm{T}(\mathfrak{n})=2+\mathfrak{n}+ 2\cdot \mathfrak{n}\cdot\ln{\mathfrak{n}} + (\sum_{\mathfrak{j}=1}^{\mathfrak{n}-1} \mathrm{T}(\mathfrak{j}))/{\mathfrak{n}} ;
\end{equation}
and for $L_1$ metric we have the following relation:
\begin{equation}\label{eq:reldiameterb}
\textstyle\mathrm{T}(\mathfrak{n})=2+\mathfrak{n}+ 2\cdot \mathfrak{n} + (\sum_{\mathfrak{j}=1}^{\mathfrak{n}-1} \mathrm{T}(\mathfrak{j}))/{\mathfrak{n}} \\
\end{equation}
We note that the worst-case complexity for this algorithm is as follows:
for Euclidean metric it is $\Theta(n^2 \cdot \log n)$ and for the $L_1$ metric
it is $\Theta(n^2)$.\qed
\end{example}

\begin{example}[Sorting with {\sc Quick-Select}]\label{ex:sortselect}
Consider a sorting algorithm
which selects the median through the {\sc Quick-Select} algorithm.
The recurrence relation is directly obtained as follows:
\begin{equation}\label{eq:relsortselect}
\textstyle\mathrm{T}(\mathfrak{n})=4+ T^*(\mathfrak{n})+\mathrm{T}\left(\lfloor{\mathfrak{n}}/{2}\rfloor\right)+\mathrm{T}\left(\lceil{\mathfrak{n}}/{2}\rceil\right)
\end{equation}
where $T^*(\centerdot)$ is an upper bound on the expected running time of {\sc Quick-Select}
(cf. Example~\ref{ex:quickselect}).
We note that the worst-case complexity for this algorithm is $\Theta(n^2)$.\qed
\end{example}

\vspace{-1.5em}
\subsection{Separable Bivariate Randomized Recurrences}\label{sec:bivariate}
\vspace{-1em}

We consider a generalization of the univariate recurrence relations to a
class of bivariate recurrence relations called \emph{separable bivariate recurrence relations}.
Similar to the univariate situation, we use `$\mathrm{T}$' to represent the (only) function call and `$\mathfrak{n}$', `$\mathfrak{m}$' to represent namely the two integer parameters.

\smallskip\noindent{\bf Separable Bivariate Recurrence Expressions.}
The syntax of \emph{separable bivariate recurrence expressions} is illustrated by $\mathfrak{e},\mathfrak{h}$ and $\mathfrak{b}$ as follows:
\begin{align*}
\mathfrak{e} & ::= \mathrm{T}\left(\mathfrak{n}, \mathfrak{m}-1\right) \mid \mathrm{T}\left(\mathfrak{n},\left\lfloor{\mathfrak{m}}/{2}\right\rfloor\right) \mid \mathrm{T}\left(\mathfrak{n},\left\lceil{\mathfrak{m}}/{2}\right\rceil\right)\\
& \mid \frac{\sum_{\mathfrak{j}=1}^{\mathfrak{m}-1} \mathrm{T}(\mathfrak{n},\mathfrak{j})}{\mathfrak{m}}
\mid \frac{1}{\mathfrak{m}}\cdot\left( \textstyle\sum_{\mathfrak{j}=\left\lceil {\mathfrak{m}}/{2}\right\rceil}^{\mathfrak{m}-1}\mathrm{T}(\mathfrak{n},\mathfrak{j})+
\textstyle\sum_{\mathfrak{j}=\left\lfloor {\mathfrak{m}}/{2}\right\rfloor}^{\mathfrak{m}-1} \mathrm{T}(\mathfrak{n},\mathfrak{j})\right)\mid c\cdot \mathfrak{e}\mid \mathfrak{e}+\mathfrak{e}
\\
\mathfrak{h} & ::= c\mid \ln{\mathfrak{n}}\mid \mathfrak{n}\mid  \mathfrak{n}\cdot\ln{\mathfrak{n}}\mid c\cdot\mathfrak{h}\mid \mathfrak{h}+\mathfrak{h}\quad\mathfrak{b}  ::= c\mid \frac{1}{\mathfrak{m}} \mid \ln{\mathfrak{m}}\mid \mathfrak{m}\mid \mathfrak{m}\cdot\ln{\mathfrak{m}}\mid c\cdot \mathfrak{b}\mid \mathfrak{b}+\mathfrak{b}
\end{align*}
The differences are that (i) we have two independent parameters $\mathfrak{n},\mathfrak{m}$, (ii) $\mathfrak{e}$ now represents an expression composed of only $\mathrm{T}$-terms, and (iii) $\mathfrak{h}$ (resp. $\mathfrak{b}$) represents arithmetic expressions for $\mathfrak{n}$ (resp. for $\mathfrak{m}$).
This class of separable bivariate recurrence expressions (often for brevity bivariate recurrence expressions)
stresses a dominant role on $\mathfrak{m}$ and a minor role on $\mathfrak{n}$, and is intended to model randomized algorithms where some parameter (to be represented by $\mathfrak{n}$) does not change value.

\smallskip\noindent{\em Substitution.}
The notion of substitution is similar to the univariate case.
Consider a function $h:\Nset\times\Nset\rightarrow\Rset$, and a bivariate recurrence expression ${\mathfrak{e}}$.
The {\em substitution function}, denoted by $\Sub({\mathfrak{e}},h)$, is the function from $\Nset\times\Nset$ into $\Rset$
such that $\Sub({\mathfrak{e}},h)(n,m)$ is the real number evaluated through substituting $h,n,m$ for
$\mathrm{T},\mathfrak{n},\mathfrak{m}$, respectively.
The substitution for $\mathfrak{h},\mathfrak{b}$ is defined in a similar way, with the difference that they both induce a univariate function.

\smallskip\noindent{\bf Bivariate recurrence relations.}
We consider {\em bivariate recurrence relations} $G=(\eql_1,\eql_2)$, which consists of
two equalities of the following form:
\begin{equation}\label{eq:birecurrel}
\eql_1: \ \mathrm{T}(\mathfrak{n},\mathfrak{m})=\mathfrak{e}+\mathfrak{h}\cdot\mathfrak{b}; \quad  \qquad \eql_2: \ \mathrm{T}(\mathfrak{n},1)=\mathfrak{h}\cdot c
\end{equation}
where $c\in(0,\infty)$ and $\mathfrak{e},\mathfrak{h},\mathfrak{b}$ are from the grammar above.

\smallskip\noindent{\em Solution to bivariate recurrence relations.}
The evaluation of bivariate recurrence relation is similar to the univariate
case.
Similar to the univariate case,
the unique solution $T_G:\Nset\times\Nset\rightarrow\Rset$ to a recurrence relation $G$ taking the form (\ref{eq:birecurrel}) is a function defined recursively as follows:
(1)~\emph{Base Step.} $T_G(n,1):=\Sub({\mathfrak{h}})(n)\cdot c$ for all $n\in\Nset$; and
(2)~\emph{Recursive Step.} $T_G(n,m):=\Sub({\mathfrak{e}},T_G)(n,m)+\Sub(\mathfrak{h})(n)\cdot\Sub(\mathfrak{b})(m)$
for all $n\in\Nset$ and $m\ge 2$.
Again the interesting algorithmic question is to reason about the infinite behaviour of $T_G$.

\vspace{-1.5em}
\subsection{Motivating Classical Examples}\label{sec:motivatingbi}
\vspace{-0.5em}
In this section we present two classical examples of randomized algorithms
where the randomized recurrence relations are bivariate.
We put the detailed illustration for this two examples in Appendix~\ref{app:motivatingbi}.


\begin{example}[{\sc Coupon-Collector}]\label{ex:coupon}
Consider the {\sc Coupon-Collector} problem~\cite[Chapter~3]{DBLP:books/cu/MotwaniR95} with $n$ different types of coupons ($n\in\Nset$).
The randomized process proceeds in rounds: at each round, a coupon is collected uniformly at random from the coupon types
the rounds continue until all the $n$ types of coupons are collected.
We model the rounds as a recurrence relation with two variables
$\mathfrak{n},\mathfrak{m}$, where $\mathfrak{n}$ represents the total number
of coupon types and $\mathfrak{m}$ represents the remaining number of uncollected
coupon types.
The recurrence relation is as follows:
\begin{equation}\label{eq:relcoupon}
\mathrm{T}(\mathfrak{n},1)=\mathfrak{n}\cdot 1; \qquad
\mathrm{T}(\mathfrak{n},\mathfrak{m})=\mathfrak{n}/{\mathfrak{m}}+ \mathrm{T}(\mathfrak{n},\mathfrak{m}-1)
\end{equation}
where $\mathrm{T}(\mathfrak{n},\mathfrak{m})$ is the expected number of rounds.
We note that the worst-case complexity for this process is $\infty$.\qed
\end{example}

\begin{example}[{\sc Channel-Conflict Resolution}]\label{ex:channel}
We consider two network scenarios in which $n$ clients are trying to get
access to a network channel.
This problem is also called the {\sc Resource-Contention Resolution}~\cite[Chapter~13]{Kleinbergbook}.
In this problem, if more than one client tries to access the channel,
then no client can access it, and if exactly one client requests access to
the channel, then the request is granted.
In the distributed setting, the clients do not share any information.
In this scenario, in each round, every client requests an access to the
channel with probability $\frac{1}{n}$.
Then for this scenario, we obtain an over-approximating recurrence relation
\begin{equation}\label{eq:relresourcea}
\mathrm{T}(\mathfrak{n},1)=\mathfrak{n}\cdot 1; \qquad
\mathrm{T}(\mathfrak{n},\mathfrak{m})=(\mathfrak{n}\cdot{e})/{\mathfrak{m}}+ \mathrm{T}(\mathfrak{n},\mathfrak{m}-1)
\end{equation}
for the expected rounds until which every client gets at least one access to the channel.
In the concurrent setting, the clients share one variable, which is the number
of clients which has not yet been granted access.
Also in this scenario, once a client gets an access the client does not
request for access again.
For this scenario, we obtain an over-approximating recurrence relation
\begin{equation}\label{eq:relresourceb}
\mathrm{T}(\mathfrak{n},1)=1\cdot 1; \qquad
\mathrm{T}(\mathfrak{n},\mathfrak{m})=1\cdot e+ \mathrm{T}(\mathfrak{n},\mathfrak{m}-1)
\end{equation}
We also note that the worst-case complexity for both the scenarios is $\infty$.\qed
\end{example}

%% file: average_case.tex
\vspace{-1.5em}
\section{Expected-Runtime Analysis}\label{sect:expruntime}
\vspace{-1em}
We focus on synthesizing logarithmic, linear, and almost-linear asymptotic bounds for recurrence relations.
Our goal is to decide and synthesize asymptotic bounds in the simple form:
$d\cdot \mathfrak{f}+\mathfrak{g}, \mathfrak{f}\in\{\ln{\mathfrak{n}},\mathfrak{n},\mathfrak{n}\cdot\ln{\mathfrak{n}}\}$ .
Informally, $\mathfrak{f}$ is the major term for time complexity, $d$ is the coefficient of $\mathfrak{f}$ to be synthesized,
and $\mathfrak{g}$ is the time complexity for the base case specified in (\ref{eq:unirecurrel}) or (\ref{eq:birecurrel}).

\smallskip\noindent{\bf Univariate Case:} The  algorithmic problem in univariate case is as follows:

\begin{compactitem}
\item {\em Input:} a univariate recurrence relation $G$ taking the form (\ref{eq:unirecurrel}) and an
expression $\mathfrak{f}\in\{\ln{\mathfrak{n}},\mathfrak{n},\mathfrak{n}\cdot\ln{\mathfrak{n}}\}$.

\item {\em Output: Decision problem.} Output ``$\mbox{\sl yes}$'' if $T_G \in \mathcal{O}(\Sub(\mathfrak{f}))$, and ``$\mbox{\sl fail}$'' otherwise.

\item {\em Output: Quantitative problem.}
A positive real number $d$ such that
\begin{equation}\label{eq:uniguess}
T_G(n) \leq  d\cdot \Sub(\mathfrak{f})(n)+c
\end{equation}
for all $n \geq 1$, 
or ``$\mbox{\sl fail}$'' otherwise, where $c$ is from (\ref{eq:unirecurrel}).
\end{compactitem}

\begin{remark}
First note that while in the problem description we consider the form $\mathfrak{f}$
part of input for simplicity, since there are only three possibilites we can simply
enumerate them, and thus have only the recurrence relation as input.
Second, in the algorithmic problem above, w.l.o.g, we consider that every
$\mathfrak{e}$ in (\ref{eq:unirecurrel}) or (\ref{eq:birecurrel}) involves at least
one $\mathrm{T}(\centerdot)$-term and one non-$\mathrm{T}(\centerdot)$-term;
this is natural since (i) for algorithms with recursion at least one $\mathrm{T}(\centerdot)$-term
should be present for the recursive call and at least one non-$\mathrm{T}(\centerdot)$-term for non-recursive base step. \qed
\end{remark}

\smallskip\noindent{\bf Bivariate Case:} The bivariate-case problem is an extension of the univariate one,
and hence the problem definitions are similar, and we present them succinctly below.
\begin{compactitem}
\item {\em Input:} a bivariate recurrence relation $G$ taking the form (\ref{eq:birecurrel}) and an expression
$\mathfrak{f}$ (similar to the univariate case).
\item {\em Output: Decision problem.} Output ``$\mbox{\sl yes}$'' if $T_G \in \mathcal{O}(\Sub(\mathfrak{f}))$, and ``$\mbox{\sl fail}$'' otherwise;
\item {\em Output: Quantitative problem.}
A positive real number $d$ such that
$T_G(n,m) \leq d\cdot \Sub(\mathfrak{f})(n,m) +c\cdot\Sub(\mathfrak{h})(n)$
for all $n,m \geq 1$, 
or ``$\mbox{\sl fail}$'' otherwise, where $c,\mathfrak{h}$ are from (\ref{eq:birecurrel}).
Note that in the expression above the term $\mathfrak{b}$ does not appear as it can be captured with
$\mathfrak{f}$ itself.
\end{compactitem}

Recall that in the above algorithmic problems obtaining the finite behaviour of
the recurrence relations is easy (through evaluation of the recurrences using
dynamic programming), and the interesting aspect is to decide the asymptotic
infinite behaviour.

%% file: synalgorithm.tex
\newcommand{\OvAp}{\mathsf{OvAp}}

\vspace{-1.5em}
\section{The Synthesis Algorithm}\label{sect:synalg}
\vspace{-1em}

In this section, we present our algorithms to synthesize
asymptotic bounds for randomized recurrence relations.

\smallskip\noindent{\em Main ideas.}
The main idea is as follows.
Consider as input a recurrence relation taking the form (\ref{eq:unirecurrel}) and an univariate recurrence expression
$\mathfrak{f}\in\{\ln{\mathfrak{n}}, \mathfrak{n},\mathfrak{n}\cdot\ln{\mathfrak{n}}\}$ which specifies the desired asymptotic bound.
We first define the standard notion of a guess-and-check function which provides a sound approach for asymptotic bound.
Based on the guess-and-check function, our algorithm executes the following steps for the univariate case.
\begin{compactenum}
\item First, the algorithm sets up a scalar variable $d$ and then constructs the template $h$ to be $n\mapsto d\cdot \Sub(\mathfrak{f})(n)+c$
for a univariate guess-and-check function.
\item Second, the algorithm computes an over-approximation $\OvAp(\mathfrak{e}, h)$ of $\Sub(\mathfrak{e}, h)$
such that the over-approximation $\OvAp(\mathfrak{e}, h)$ will involve terms from $\mathfrak{n}^k,\ln^\ell{\mathfrak{n}}$ (for $k,\ell\in\Nset_0$) only.
Note that $k,\ell$ may be greater than $1$, so the above expressions are not necessarily linear (they can be quadratic or cubic for example).

\item Finally, the algorithm synthesizes a value for $d$
such that $\OvAp(\mathfrak{e},h)(n)\le h(n)$ for all $n\ge 2$ through truncation
of $[2,\infty)\cap\Nset$ into a finite range and a limit behaviour analysis (towards $\infty$).
\end{compactenum}
Our algorithm for bivariate cases is a reduction to the univariate case.

\smallskip\noindent{\bf Guess-and-Check functions.}
We follow the standard guess-and-check technique to solve simple recurrence relations.
Below we first fix a univariate recurrence relation $G$ taking the form (\ref{eq:unirecurrel}).
By an easy induction on $n$ (starting from the $N$ specified in Definition~\ref{def:uniguess})
we obtain Theorem~\ref{thm:uniguess}.

\begin{definition}[Univariate Guess-and-Check Functions]\label{def:uniguess}
Let $G$ be a univariate recurrence relation taking the form (\ref{eq:unirecurrel}).
A function $h:\Nset\rightarrow\Rset$ is a \emph{guess-and-check} function for $G$ if there exists a natural number $N\in\Nset$ such that:
(1) {\em (Base Condition)} $T_G(n)\le h(n)$ for all $1\le n\le N$, and
(2) {\em (Inductive Argument)} $\Sub(\mathfrak{e},h) (n)\le h(n)$ for all $n> N$.
\end{definition}

\begin{theorem}[Guess-and-Check, Univariate Case]\label{thm:uniguess}
If a function $h:\Nset\rightarrow\Rset$ is a \emph{guess-and-check} function for a univariate recurrence relation $G$ taking the form (\ref{eq:unirecurrel}), then $T_G(n)\le h(n)$ for all $n\in\Nset$.
\end{theorem}

We do not explicitly present the definition for guess-and-check functions in the bivariate case,
since we will present a reduction of the analysis of separable bivariate recurrence relations to that
of the univariate ones (cf. Section~\ref{sect:bisynth}).

\smallskip\noindent{\bf Overapproximations for Recurrence Expressions.}
We now develop tight overapproximations for logarithmic terms.
In principle, we use Taylor's Theorem to approximate logarithmic terms such as $\ln{(n-1)},\ln{\lfloor\frac{n}{2}\rfloor}$,
and integral to approximate summations of logarithmic terms.
All the results below are technical and depends on basic calculus
(the detailed proofs are in the Appendix~\ref{app:overapprox}).

\begin{proposition}\label{prop:lnflooroverapprox}
For all natural number $n\ge 2$:
\[
(1)\ \ln{n}-\ln{2}-\frac{1}{n-1}\le \ln{\left\lfloor \frac{n}{2}\right\rfloor}\le \ln{n}-\ln{2};
(2)\ \ln{n}-\ln{2}\le \ln{\left\lceil \frac{n}{2}\right\rceil}\le \ln{n}-\ln{2}+\frac{1}{n}~~.
\]
\end{proposition}

\begin{proposition}\label{prop:nminusoneoverapprox}
For all natural number $n\ge 2$:
$\ln{n}-\frac{1}{n-1}\le\ln{(n-1)}\le \ln{n}-\frac{1}{n}$~~.
\end{proposition}

\begin{proposition}\label{prop:integralapproximation}
For all natural number $n\geq 2$:
\begin{compactitem}
\item $\int_1^n \frac{1}{x}\,\mathrm{d}x-\sum_{j=1}^{n-1} \frac{1}{j}\in \left[-0.7552,-\frac{1}{6}\right]$;
\item $\int_1^n \ln{x}\,\mathrm{d}x-\left(\sum_{j=1}^{n-1} \ln{j}\right) - \frac{1}{2}\cdot \int_1^n \frac{1}{x}\,\mathrm{d}x\in \left[-\frac{1}{12}, 0.2701\right]$;
\item $\int_1^n x\cdot \ln{x}\,\mathrm{d}x-\left(\sum_{j=1}^{n-1} j\cdot\ln{j}\right)-\frac{1}{2}\cdot\int_1^n \ln{x}\,\mathrm{d}x+\frac{1}{12}\cdot \int_1^n \frac{1}{x}\,\mathrm{d}x-\frac{n-1}{2}\in \left[-\frac{19}{72},0.1575\right]$.
\end{compactitem}
\end{proposition}

Note that Proposition~\ref{prop:integralapproximation} is non-trivial since it approximates summation of reciprocal and logarithmic terms up to a constant deviation.
For example, one may approximate $\sum_{j=1}^{n-1} \ln{j}$ directly by $\int_1^n \ln{x}\,\mathrm{d}x$, but this approximation deviates up to a logarithmic term from Proposition~\ref{prop:integralapproximation}.
From Proposition~\ref{prop:integralapproximation}, we establish a tight approximation for summation of logarithmic or reciprocal terms.


\begin{example}\label{ex:overapprox}
Consider the summation $\sum_{j=\left\lceil\frac{n}{2}\right\rceil}^{n-1}\ln{j}+ \sum_{j=\left\lfloor\frac{n}{2}\right\rfloor}^{n-1} \ln{j}\quad (n\ge 4)$.
By Proposition~\ref{prop:integralapproximation}, we can over-approximate it as
\[
2\cdot\left(\Gamma_{\ln{\mathfrak{n}}}\left(n\right)+\frac{1}{12}\right)
-\left(\Gamma_{\ln{\mathfrak{n}}}\left(\left\lceil\frac{n}{2}\right\rceil\right)+\Gamma_{\ln{\mathfrak{n}}}\left(\left\lfloor\frac{n}{2}\right\rfloor\right)-0.5402\right)
\]
where $\Gamma_{\ln{\mathfrak{n}}}(n)  :=
\int_1^n\ln{x}\,\mathrm{d}x-\frac{1}{2}\cdot\int_1^n\frac{1}{x}\,\mathrm{d}x
 =
n\cdot\ln{n}-n-\frac{\ln{n}}{2}+1$.
By using Proposition~\ref{prop:lnflooroverapprox}, the above expression is roughly
$n\cdot\ln{n}-(1-\ln{2})\cdot n+\frac{1}{2}\cdot\ln{n}+0.6672+\frac{1}{2\cdot n}$
(for details see Appendix~\ref{app:overapprox}).\qed
\end{example}

\begin{remark}\label{rmk:exthigerdegree}
Although we do approximation for terms related to only almost-linear bounds, Proposition~\ref{prop:integralapproximation} can be extended to logarithmic bounds with higher degree (e.g., $n^3\ln n$) since integration of such bounds can be obtained in closed forms.\qed
\end{remark}

\vspace{-1.5em}
\subsection{Algorithm for Univariate Recurrence Relations}\label{sect:unisynth}
\vspace{-1em}
We present our algorithm to synthesize a guess-and-check function in form~(\ref{eq:uniguess})
for univariate recurrence relations.
We present our algorithm in two steps. First, we present the decision version,
and then we present the quantitative version that synthesizes the associated constant.
The two key aspects are over-approximation and use of pseudo-polynomials,
and we start with over-approximation.
We relegate some technical details to Appendix~\ref{app:unisynth}.

\begin{definition}[Overapproximation]\label{def:unioverapprox}
Let $\mathfrak{f}\in\{\ln{\mathfrak{n}},\mathfrak{n},\mathfrak{n}\cdot\ln{\mathfrak{n}}\}$.
Consider a univariate recurrence expression $\mathfrak{g}$, constants $d$ and $c$,
and the function $h= d \cdot \Sub(\mathfrak{f}) + c$.
We define the {\em over-approximation function}, denoted $\OvAp(\mathfrak{g},h)$,
recursively as follows.
\begin{itemize}
\item {\em Base Step A.} If $\mathfrak{g}$ is one of the following:
$c', \mathfrak{n}, \ln{\mathfrak{n}}, \mathfrak{n}\cdot \ln{\mathfrak{n}},\frac{1}{\mathfrak{n}}$,
then $\OvAp(\mathfrak{g},h):=\Sub({\mathfrak{g}})$.

\item {\em Base Step B.} If $\mathfrak{g}$ is a single term which involves $\mathrm{T}$,
then we define $\OvAp(\mathfrak{g},h)$
from over-approximations 
Proposition~\ref{prop:lnflooroverapprox}--~\ref{prop:integralapproximation}.
In details, $\OvAp(\mathfrak{g},h)$
is obtained from $\Sub(\mathfrak{g},h)$ by first over-approximating any summation through Proposition~\ref{prop:integralapproximation}
(i.e., through those $\Gamma_{(\centerdot)}$ functions defined below Proposition~\ref{prop:integralapproximation}), then over-approximating any
$\ln{(\mathfrak{n}-1)}, \left\lfloor\frac{\mathfrak{n}}{2}\right\rfloor, \left\lceil \frac{\mathfrak{n}}{2}\right\rceil, \ln{\left\lfloor\frac{\mathfrak{n}}{2}\right\rfloor}, \ln{\left\lceil \frac{\mathfrak{n}}{2}\right\rceil}$
by Proposition~\ref{prop:lnflooroverapprox} and Proposition~\ref{prop:nminusoneoverapprox}.
The details of the important over-approximations are illustrated explicitly
in Table~\ref{tbl:unioverapprox}.

\item {\em Recursive Step.} We have two cases:
(a)~If $\mathfrak{g}$ is $\mathfrak{g}_1+\mathfrak{g}_2$,
then $\OvAp(\mathfrak{g},h)$ is $\OvAp(\mathfrak{g}_1,h)+\OvAp(\mathfrak{g}_2,h)$.
(b)~If $\mathfrak{g}$ is $c'\cdot\mathfrak{g}'$, then
$\OvAp(\mathfrak{g},h)$ is $c'\cdot\OvAp(\mathfrak{g}',h)$.
\end{itemize}
\end{definition}

\begin{table}
\caption{Illustration for Definition~\ref{def:unioverapprox} where the notations are given in the top-left corner.}
\label{tbl:unioverapprox}
\centering
\scalebox{0.82}{
\begin{tabular}{|c|c|c|c|}
\hline
Notation & Expression & $\mathfrak{f}$, $\mathrm{T}$-term  &  Over-approximation \\
\hline
$\mathfrak{e}_1$ & $\mathrm{T}(\mathfrak{n}-1)$ & $\ln{\mathfrak{n}}$, $\mathfrak{e}_1$ &
$\ln{\mathfrak{n}}-\frac{1}{\mathfrak{n}}$\\
\hline
$\mathfrak{e}_2$ & $\mathrm{T}\left(\left\lfloor\frac{\mathfrak{n}}{2}\right\rfloor\right)$ & $\ln{\mathfrak{n}}$, $\mathfrak{e}_2$ & $\ln{\mathfrak{n}}-\ln{2}$
\\
\hline
$\mathfrak{e}_3$ & $\mathrm{T}\left(\left\lceil\frac{\mathfrak{n}}{2}\right\rceil\right)$ & $\ln{\mathfrak{n}}$, $\mathfrak{e}_3$ & $\ln{\mathfrak{n}}-\ln{2}+\frac{1}{\mathfrak{n}}$ \\
\hline
$\mathfrak{e}_4$ & $\frac{1}{\mathfrak{n}}\cdot \sum_{\mathfrak{j}=1}^{\mathfrak{n}-1} \mathrm{T}(\mathfrak{j})$ & $\ln{\mathfrak{n}}$, $\mathfrak{e}_4$ & $\ln{\mathfrak{n}}-1-\frac{\ln{\mathfrak{n}}}{2\cdot\mathfrak{n}} +\frac{13}{12}\cdot\frac{1}{\mathfrak{n}}$\\
\hline
$\mathfrak{e}_5$ & $\frac{1}{\mathfrak{n}}\cdot\left(\sum_{\mathfrak{j}=\left\lceil\frac{\mathfrak{n}}{2}\right\rceil}^{\mathfrak{n}-1}\mathrm{T}(\mathfrak{j})+ \sum_{\mathfrak{j}=\left\lfloor\frac{\mathfrak{n}}{2}\right\rfloor}^{\mathfrak{n}-1} \mathrm{T}(\mathfrak{j})\right)$ &
$\ln{\mathfrak{n}}$, $\mathfrak{e}_5$ & $\ln{\mathfrak{n}}-(1-\ln{2})+\frac{\ln{\mathfrak{n}}}{2\cdot \mathfrak{n}}+\frac{0.6672}{\mathfrak{n}}+\frac{1}{2\cdot \mathfrak{n}^2} $ \\
\hline
\hline
$\mathfrak{f}$, $\mathrm{T}$-term &  Over-approximation & $\mathfrak{f}$, $\mathrm{T}$-term  &  Over-approximation \\
\hline
$\mathfrak{n}$, $\mathfrak{e}_1$ & $\mathfrak{n}-1$ & $\mathfrak{n}\cdot\ln{\mathfrak{n}}$, $\mathfrak{e}_1$ & $\mathfrak{n}\cdot\ln{\mathfrak{n}}-\ln{\mathfrak{n}}-1+\frac{1}{\mathfrak{n}}$ \\
\hline
$\mathfrak{n}$, $\mathfrak{e}_2$ & $\frac{\mathfrak{n}}{2}$ & $\mathfrak{n}\cdot\ln{\mathfrak{n}}$, $\mathfrak{e}_2$ & $\frac{1}{2}\cdot\mathfrak{n}\cdot \ln{\mathfrak{n}}-\frac{\ln{2}}{2}\cdot\mathfrak{n}$ \\
\hline
$\mathfrak{n}$, $\mathfrak{e}_3$ & $\frac{\mathfrak{n}+1}{2}$ & $\mathfrak{n}\cdot\ln{\mathfrak{n}}$, $\mathfrak{e}_3$ & $\frac{\mathfrak{n}\cdot \ln{\mathfrak{n}}}{2}-\frac{\ln{2}}{2}\cdot \mathfrak{n}+\frac{1-\ln{2}}{2}+\frac{\ln{\mathfrak{n}}}{2}+\frac{1}{2\cdot\mathfrak{n}}$ \\
\hline
$\mathfrak{n}$, $\mathfrak{e}_4$ & $\frac{\mathfrak{n}-1}{2}$ & $\mathfrak{n}\cdot\ln{\mathfrak{n}}$, $\mathfrak{e}_4$ & $\frac{\mathfrak{n}\cdot\ln{\mathfrak{n}}}{2}-\frac{\mathfrak{n}}{4}-\frac{\ln{\mathfrak{n}}}{2}+\frac{\ln{\mathfrak{n}}}{12\cdot\mathfrak{n}}+\frac{0.5139}{\mathfrak{n}}$ \\
\hline
\multirow{2}{*}{$\mathfrak{n}$, $\mathfrak{e}_5$} & \multirow{2}{*}{$\frac{3}{4}\cdot \mathfrak{n}-\frac{1}{4\cdot \mathfrak{n}}$} & \multirow{2}{*}{$\mathfrak{n}\cdot\ln{\mathfrak{n}}$, $\mathfrak{e}_5$} &  $\frac{3}{4}\cdot \mathfrak{n}\cdot \ln{\mathfrak{n}}-0.2017\cdot \mathfrak{n}-\frac{1}{2}\cdot \ln{\mathfrak{n}}$ \\
 & & & $-0.2698+\frac{\ln{\mathfrak{n}}}{8\cdot\mathfrak{n}}+\frac{1.6369}{\mathfrak{n}}+\frac{1}{2\cdot\mathfrak{n}\cdot(\mathfrak{n}-1)}+\frac{1}{4\cdot \mathfrak{n}^2}$ \\
 \hline
\end{tabular}
}
\vspace{-1em}
\end{table}

\begin{example}\label{ex:reloverapprox}
Consider the recurrence relation for Sherwood's {\sc Randomized-Search} (cf.~(\ref{eq:relrandsearch})).
Choose $\mathfrak{f}=\ln{\mathfrak{n}}$ and then the template $h$ becomes
$n\mapsto d\cdot \ln{n}+1$.
From Example~\ref{ex:overapprox}, we have that the over-approximation for
$6+\frac{1}{\mathfrak{n}}\cdot\left( \sum_{\mathfrak{j}=\left\lceil\frac{\mathfrak{n}}{2}\right\rceil}^{\mathfrak{n}-1}\mathrm{T}(\mathfrak{j})+ \sum_{\mathfrak{j}=\left\lfloor\frac{\mathfrak{\mathfrak{n}}}{2}\right\rfloor}^{\mathfrak{\mathfrak{n}}-1} \mathrm{T}(\mathfrak{j})\right)$
when $n\ge 4$ is
$7+ d\cdot \left[\ln{n}-(1-\ln{2})+\frac{\ln{n}}{2\cdot n}+\frac{0.6672}{n}+\frac{1}{2\cdot n^2}\right]$
(the second summand comes from an over-approximation of
$\frac{1}{\mathfrak{n}}\cdot\left( \sum_{\mathfrak{j}=\left\lceil\frac{\mathfrak{n}}{2}\right\rceil}^{\mathfrak{n}-1}d\cdot \ln{\mathfrak{j}}+ \sum_{\mathfrak{j}=\left\lfloor\frac{\mathfrak{\mathfrak{n}}}{2}\right\rfloor}^{\mathfrak{\mathfrak{n}}-1} d\cdot \ln{\mathfrak{j}}\right)$).\qed
\end{example}

\begin{remark}
Since integrations of the form $\int x^k\ln^l x\,\mathrm{d}x$ can be calculated in closed forms (cf. Remark~\ref{rmk:exthigerdegree}),  Table~\ref{tbl:unioverapprox} can be extended to logarithmic expressions with higher order, e.g., $\mathfrak{n}^2\ln \mathfrak{n}$.\qed
\end{remark}

\smallskip\noindent{\em Pseudo-polynomials.}
Our next step is to define the notion of (univariate) pseudo-polynomials
which extends normal polynomials with logarithm.
This notion is crucial to handle inductive arguments in the definition
of (univariate) guess-and-check functions.

\begin{definition}[Univariate Pseudo-polynomials]
A univariate pseudo-polynomial (w.r.t logarithm) is a function $p:\Nset\rightarrow\Rset$ such that there exist non-negative integers $k,\ell \in \Nset_0$ and real numbers $a_i,b_i$'s such that for all $n\in\Nset$,
\vspace{-0.5em}
\begin{equation}\label{eq:pseudopoly}
\textstyle p(n)=\sum_{i=0}^{k} a_i\cdot n^{i}\cdot \ln{n}+\sum_{i=0}^{\ell} b_i\cdot n^{i}.
\end{equation}
\vspace{-1em}
\end{definition}
W.l.o.g, we consider that in the form (\ref{eq:pseudopoly}), it holds that
(i) $a^2_k+b^2_\ell\ne 0$,
(ii) either $a_k\ne 0$ or $k=0$, and (iii) similarly either $b_\ell\ne 0$ or $\ell=0$.

\smallskip\noindent{\em Degree of pseudo-polynomials.}
Given a univariate pseudo-polynomial $p$ in the form (\ref{eq:pseudopoly}), we define the \emph{degree} $\mathrm{deg}(p)$ of $p$ by:
$\mathrm{deg}(p)= k+\frac{1}{2}$  if $k\ge \ell\mbox{ and }a_k\ne 0$ and $\ell$ otherwise.
Intuitively, if the term with highest degree involves logarithm, then we increase the degree by $1/2$,
else it is the power of the highest degree term.

\smallskip\noindent{\em Leading term $\overline{p}$.}
The \emph{leading term} $\overline{p}$ of a pseudo-polynomial $p$ in the form (\ref{eq:pseudopoly}) is a function
$\overline{p}:\Nset\rightarrow\Rset$ defined as follows:
$\overline{p}(n)=a_{k}\cdot n^{k}\cdot \ln{n}  \mbox{ if }k\ge \ell\mbox{ and }a_k\ne 0$; and
$b_{\ell}\cdot n^{\ell} \mbox{ otherwise }$; for all $n\in\Nset$.
Furthermore, we define $C_p$ to be the (only) coefficient of $\overline{p}$.

With the notion of pseudo-polynomials, the inductive argument of guess-and-check functions can be soundly transformed into an inequality between pseudo-polynomials.

\begin{lemma}\label{lemm:unitrans}
Let $\mathfrak{f}\in\{\ln{\mathfrak{n}},\mathfrak{n},\mathfrak{n}\cdot\ln{\mathfrak{n}}\}$
and $c$ be a constant.
For all univariate recurrence expressions $\mathfrak{g}$, there exists pseudo-polynomials
$p$ and $q$ such that coefficients (i.e., $a_i,b_i$'s in~(\ref{eq:pseudopoly}))
of $q$ are all non-negative, $C_q>0$ and the following assertion holds:
for all $d>0$ and for all $n\ge 2$, with $h=d\cdot \Sub({\mathfrak{f}})+c$,
the inequality $\OvAp(\mathfrak{g}, h)(n)\le h(n)$ is equivalent to $d\cdot p(n)\ge q(n)$.
\end{lemma}

\begin{remark}\label{rem:unitrans}
In the above lemma, though we only refer to existence of pseudo-polynomials $p$ and $q$,
they can actually be computed in linear time, because $p$ and $q$ are obtained by simple rearrangements
of terms from $\OvAp(\mathfrak{g}, h)$ and $h$, respectively.
\end{remark}

\begin{example}\label{ex:inequality}
Let us continue with Sherwood's {\sc Randomized-Search}.
Again choose $h=d\cdot\ln{\mathfrak{n}}+1$.
From Example~\ref{ex:reloverapprox}, we obtain that for every $n\ge 4$, the inequality
\begin{align*}
d\cdot\ln{n}+1\ge  7+ d\cdot \left[\ln{n}-(1-\ln{2})+\frac{\ln{n}}{2\cdot n}+\frac{0.6672}{n}+\frac{1}{2\cdot n^2}\right]
\end{align*}
resulting from over-approximation and the inductive argument of guess-and-check functions
 is equivalent to
$d\cdot\left[(1-\ln{2})\cdot n^2-\frac{n\cdot\ln{n}}{2}-0.6672\cdot n-\frac{1}{2}\right]\ge  6\cdot n^2$.\qed
\end{example}

As is indicated in Definition~\ref{def:uniguess},
our aim is to check whether $ \OvAp(\mathfrak{g}, h)(n)\le h(n)$
holds for sufficiently large $n$.
The following proposition provides a sufficient and necessary condition for checking whether
$d\cdot p(n)\ge q(n)$ holds for sufficiently large $n$.

\begin{proposition}\label{prop:unisufflarge}
Let $p,q$ be pseudo-polynomials such that $C_q>0$ and
all coefficients of $q$ are non-negative.
Then there exists a real number $d>0$ such that
$d\cdot p(n)\ge q(n)$
for sufficiently large $n$ iff $\mathrm{deg}(p)\ge \mathrm{deg}(q)$ and $C_p>0$.
\end{proposition}

Note that by Definition~\ref{def:uniguess} and the special form (\ref{eq:uniguess}) for univariate guess-and-check functions, a function in form (\ref{eq:uniguess}) needs only
to satisfy the inductive argument in order to be a univariate guess-and-check function: once a value for $d$ is synthesized for a sufficiently large $N$, one can scale the value so that the base condition
is also satisfied.
Thus from the sufficiency of Proposition~\ref{prop:unisufflarge}, our decision algorithm that checks the existence of some guess-and-check function in form (\ref{eq:uniguess}) is presented below.
Below we fix an input univariate recurrence relation $G$ taking the form (\ref{eq:unirecurrel}) and an input expression $\mathfrak{f}\in\{\ln{\mathfrak{n}},\mathfrak{n},\mathfrak{n}\cdot\ln{\mathfrak{n}}\}$.

\textbf{Algorithm} $\mbox{\sl UniDec}$: Our algorithm, namely $\mbox{\sl UniDec}$, for the decision problem of
the univariate case, has the following steps.
\begin{compactenum}
\item {\em Template.} The algorithm establishes a scalar variable $d$ and sets up the template
$d\cdot \mathfrak{f}+c$ for a univariate guess-and-check function.
\item {\em Over-approximation.}
Let $h$ denote $d \cdot \Sub(\mathfrak{f}) +c$. The algorithm calculates the over-approximation function $\OvAp(\mathfrak{e},h)$,
where $\mathfrak{e}$ is from (\ref{eq:unirecurrel}).
\item {\em Transformation.} The algorithm transforms the inequality
$\OvAp(\mathfrak{e},h)(n) \le h(n) ~~(n\in\Nset)$
for inductive argument of guess-and-check functions through Lemma~\ref{lemm:unitrans} equivalently into
$d\cdot p(n)\ge q(n)~~(n\in\Nset)$,
where $p,q$ are pseudo-polynomials obtained in linear-time through rearrangement of terms from
$\OvAp(\mathfrak{e},h)$ and $h$ (see Remark~\ref{rem:unitrans}).

\item {\em Coefficient Checking.} The algorithm examines cases on $C_p$. If $C_p> 0$ and $\mathrm{deg}(p)
\ge \mathrm{deg}(q)$,
then  algorithm outputs ``$\mbox{\sl yes}$'' meaning that ``there exists a univariate guess-and-check function'';
otherwise, the algorithm outputs ``$\mbox{\sl fail}$''.
\end{compactenum}

\begin{theorem}[Soundness for $\mbox{\sl UniDec}$]\label{thm:soundnessunidec}
If $\mbox{\sl UniDec}$ outputs ``$\mbox{\sl yes}$'', then there exists a univariate guess-and-check function in form~(\ref{eq:uniguess})
for the inputs $G$ and $\mathfrak{f}$.
The algorithm is a linear-time algorithm in the size of the input recurrence relation.
\end{theorem}

\begin{example}
Consider Sherwood's {\sc Randomized-Search} recurrence relation (cf.~(\ref{eq:relrandsearch})) and $\mathfrak{f}=\ln{\mathfrak{n}}$ as the input.
As illustrated in Example~\ref{ex:reloverapprox} and Example~\ref{ex:inequality}, the algorithm
asserts that the asymptotic behaviour is $\mathcal{O}(\ln{n})$.\qed
\end{example}

\begin{remark}
From the tightness of our over-approximation (up to only constant deviation) and
the sufficiency and necessity of Proposition~\ref{prop:unisufflarge},
the $\mbox{\sl UniDec}$ algorithm can handle a large class of univariate recurrence relations.
Moreover, the algorithm is quite simple and efficient (linear-time).
However, we do not know whether our approach is complete. We suspect that there is certain intricate recurrence relations that will make our approach fail.
\end{remark}

\noindent{\bf Analysis of examples of Section~\ref{sec:motivatinguni}.}
Our algorithm can decide the following optimal bounds for the examples of Section~\ref{sec:motivatinguni}.
\begin{compactenum}
\item For Example~\ref{ex:randsearch} we obtain an $\mathcal{O}(\log n)$ bound (recall worst-case bound is $\Theta(n)$).
\item For Example~\ref{ex:quicksort} we obtain an $\mathcal{O}(n\cdot\log n)$ bound (recall worst-case bound is $\Theta(n^2)$).
\item For Example~\ref{ex:quickselect} we obtain an $\mathcal{O}(n)$ bound (recall worst-case bound is $\Theta(n^2)$).
\item For Example~\ref{ex:diameter} we obtain an $\mathcal{O}(n\cdot\log n)$ (resp. $\mathcal{O}(n)$) bound for Euclidean metric
(resp. for $L_1$ metric), whereas the worst-case bound is $\Theta(n^2\cdot\log n)$ (resp. $\Theta(n^2)$).
\item For Example~\ref{ex:sortselect} we obtain an $\mathcal{O}(n\cdot\log n)$ bound (recall worst-case bound is $\Theta(n^2)$).
\end{compactenum}
In all cases above, our algorithm decides the asymptotically optimal bounds for the expected-runtime
analysis, whereas the worst-case analysis grossly over-estimate the expected-runtime bounds.

\smallskip\noindent{\bf Quantitative bounds.}
Above we have already established that our linear-time decision algorithm can establish
the asymptotically optimal bounds for the recurrence relations of several classical algorithms.
We now take the next step to obtain even explicit quantitative bounds, i.e., to synthesize
the associated constants with the asymptotic complexity.
To tackle these situations, we derive a following proposition which gives explicitly a
threshold for ``sufficiently large numbers''.
We first explicitly constructs a threshold for ``sufficiently large numbers''.
Then we show in Proposition~\ref{prop:unisufflargeN}
that $N_{\epsilon,p,q}$ is indeed what we need.

\begin{definition}[Threshold $N_{\epsilon,p,q}$ for Sufficiently Large Numbers]\label{def:unisuffN}
Let $p,q$ be two univariate pseudo-polynomials
$p(n)=\sum_{i=0}^{k} a_i\cdot n^{i}\cdot \ln{n}+\sum_{i=0}^{\ell} b_i\cdot n^{i}$~,
$q(n)=\sum_{i=0}^{k'} a'_i\cdot n^{i}\cdot \ln{n}+\sum_{i=0}^{\ell'} b'_i\cdot n^{i}$
such that $\mathrm{deg}(p)\ge \mathrm{deg}(q)$ and $C_p,C_q>0$. Then given any $\epsilon\in (0,1)$,
the number $N_{\epsilon,p,q}$ is defined as the smallest natural number such that both $x,y$ (defined below) is smaller than $\epsilon$:
\begin{compactitem}
\item $x=-1+\sum_{i=0}^{k} |a_i|\cdot \frac{N^{i}\cdot \ln{N}}{\overline{p}(N)}+\sum_{i=0}^{\ell} |b_i|\cdot \frac{N^{i}}{\overline{p}(N)}$ ;
\item $y=-\mathbf{1}_{\mathrm{deg}(p)=\mathrm{deg}(q)}\cdot\frac{C_q}{C_p}+\sum_{i=0}^{k'} |a'_i|\cdot \frac{N^{i}\cdot \ln{N}}{\overline{p}(N)}+\sum_{i=0}^{\ell'} |b'_i|\cdot \frac{N^{i}}{\overline{p}(N)}$ .
\end{compactitem}
where $\mathbf{1}_{\mathrm{deg}(p)=\mathrm{deg}(q)}$ equals $1$ when ${\mathrm{deg}(p)=\mathrm{deg}(q)}$ and $0$ otherwise.
\end{definition}

\begin{proposition}\label{prop:unisufflargeN}
Consider two univariate pseudo-polynomials $p,q$ such that $\mathrm{deg}(p)\ge \mathrm{deg}(q)$, all coefficients of $q$ are non-negative and $C_p,C_q>0$.
Then given any $\epsilon\in (0,1)$,
$\frac{q(n)}{p(n)}\le \frac{\mathbf{1}_{\mathrm{deg}(p)=\mathrm{deg}(q)}\cdot \frac{C_q}{C_p}+\epsilon}{1-\epsilon}$
for all $n\ge N_{\epsilon,p,q}$ (for $N_{\epsilon,p,q}$ of Definition~\ref{def:unisuffN}).
\end{proposition}

With Proposition~\ref{prop:unisufflargeN}, we describe our algorithm $\mbox{\sl UniSynth}$ which outputs explicitly a value for $d$ (in (\ref{eq:uniguess})) if $\mbox{\sl UniDec}$ outputs yes.
Below we fix an input univariate recurrence relation $G$ taking the form (\ref{eq:unirecurrel}) and an input expression $\mathfrak{f}\in\{\ln{\mathfrak{n}},\mathfrak{n},\mathfrak{n}\cdot\ln{\mathfrak{n}}\}$.
Moreover, the algorithm takes $\epsilon>0$ as another input, which is basically a parameter to choose the threshold for finite behaviour.
For example, smaller $\epsilon$ leads to large threshold, and vice-versa.
Thus we provide a flexible algorithm as the threshold can be varied with the choice of $\epsilon$.

\textbf{Algorithm} $\mbox{\sl UniSynth}$: Our algorithm for the quantitative problem has the following steps:
\begin{compactenum}
\item {\em Calling $\mbox{\sl UniDec}$.} The algorithm calls $\mbox{\sl UniDec}$, and if it
returns ``$\mbox{\sl fail}$'', then return ``$\mbox{\sl fail}$'', otherwise execute the following steps.
Obtain the following inequality $d\cdot p(n)\ge q(n)~~(n\in\Nset)$
from the transformation step of $\mbox{\sl UniDec}$.
\item {\em Variable Solving.}
The algorithm calculates $N_{\epsilon, p,q}$ for a given $\epsilon\in(0,1)$ by e.g. repeatedly increasing $n$ (see Definition~\ref{def:unisuffN})
and outputs the value of $d$ as the least number such that
the following two conditions hold:
(i)~for all $2\le n< N_{\epsilon, p,q}$, we have $\Eval(G)(n)\le d\cdot \Sub({\mathfrak{f}})(n)+c$
(recall $\Eval(G)(n)$ can be computed in linear time), and
(ii)~we have $d\ge \frac{\mathbf{1}_{\mathrm{deg}(p)=\mathrm{deg}(q)}\cdot \frac{C_q}{C_p}+\epsilon}{1-\epsilon}$.
\end{compactenum}

\begin{theorem}[Soundness for $\mbox{\sl UniSynth}$]\label{thm:soundnessunisynth}
If the algorithm $\mbox{\sl UniSynth}$ outputs a real number $d$, then $d\cdot \Sub(\mathfrak{f})+c$ is a univariate guess-and-check function for $G$.
\end{theorem}


\begin{example}
Consider the recurrence relation for Sherwood's {\sc Randomized-Search} (cf.~(\ref{eq:relrandsearch})) and $\mathfrak{f}=\ln{\mathfrak{n}}$.
Consider that $\epsilon:=0.9$.
From Example~\ref{ex:reloverapprox} and Example~\ref{ex:inequality}, the algorithm establishes the inequality
$d\ge \frac{ 6}{(1-\ln{2})-\frac{\ln{n}}{2\cdot n}-\frac{0.6672}{n}-\frac{1}{2\cdot n^2}}$
and finds that $N_{0.9,p,q}=6$.
Then the algorithm finds $d=204.5335$ through the followings:
(a) $\Eval(G)(2)=7\le d\cdot \ln{2}+1$;
(b) $\Eval(G)(3)=11\le d\cdot \ln{3}+1$;
(c) $\Eval(G)(4)=15\le d\cdot \ln{4}+1$;
(d) $\Eval(G)(5)=17.8\le d\cdot \ln{5}+1$;
(e) $d\ge \frac{\frac{6}{1-\ln{2}}+0.9}{1-0.9}$.
Thus, by Theorem~\ref{thm:uniguess}, the expected running time of the algorithm
has an upper bound $204.5335\cdot \ln{n}+1$.
Later in Section~\ref{sect:experiments}, we show that one can obtain a much better $d=19.762$ through our algorithms  by choosing $\epsilon:=0.01$, which is quite good since the optimal value lies in $[15.129, 19.762]$ (cf. the first item {\sc R.-Sear.} in Table~\ref{tab:experiments}).\qed
\end{example}

\vspace{-1.5em}
\subsection{Algorithm for Bivariate Recurrence Relations}\label{sect:bisynth}
\vspace{-1em}

In this part, we present our results for the separable bivariate recurrence relations.
The key idea is to use separability to reduce the problem to univariate recurrence relations.
There are two key steps which we describe below.

\smallskip\noindent{\em Step~1.}
The first step is to reduce a separable bivariate recurrence relation to a univariate one.

\begin{definition}[From $G$ to $\Uni{G}$]
Let $G$ be a separable bivariate recurrence relation taking the form~(\ref{eq:birecurrel}).
The univariate recurrence relation $\Uni{G}$ from $G$ is defined by eliminating any occurrence of $\mathfrak{n}$ and replacing any occurrence of $\mathfrak{h}$ with $1$.
\end{definition}
Informally, $\Uni{G}$ is obtained from $G$ by simply eliminating the roles of $\mathfrak{h},\mathfrak{n}$.
The following example illustrates the situation for {\sc Coupon-Collector} example.

\begin{example}
Consider $G$ to be the recurrence relation (\ref{eq:relcoupon}) for {\sc Coupon-Collector} example.
Then $\Uni{G}$ is as follows:
$\mathrm{T}(\mathfrak{n})=\frac{1}{\mathfrak{n}}+ \mathrm{T}(\mathfrak{n}-1)$ and $\mathrm{T}(1)=1$.
\qed
\end{example}

\smallskip\noindent{\em Step~2.}
The second step is to establish the relationship between $T_G$ and $T_{\Uni{G}}$, which is handled by the following proposition,
whose proof is an easy induction on $m$.

\begin{proposition}\label{prop:reduction}
For any separable bivariate recurrence relation $G$ taking the form (\ref{eq:birecurrel}), the solution $T_G$ is equal to $(n,m)\mapsto \Sub(\mathfrak{h})(n) \cdot T_{\Uni{G}}(m)$.
\end{proposition}

\smallskip {\em Description of the Algorithm.}
With Proposition~\ref{prop:reduction}, the algorithm for separable bivariate recurrence relations is straightforward:
simply compute $\Uni{G}$ for $G$ and then call the algorithms for univariate case presented in Section~\ref{sect:unisynth}.

\smallskip\noindent{\bf Analysis of examples in Section~\ref{sec:motivatingbi}.}
Our algorithm can decide the following optimal bounds for the examples of Section~\ref{sec:motivatingbi}.
\begin{compactenum}
\item For Example~\ref{ex:coupon} we obtain an $\mathcal{O}(n\cdot \log m)$ bound, whereas the worst-case bound is $\infty$.

\item For Example~\ref{ex:channel} we obtain an $\mathcal{O}(n\cdot\log m)$ bound for distributed setting and $\mathcal{O}(m)$ bound for concurrent setting, whereas the worst-case bounds are both $\infty$.
\end{compactenum}
Note that for all our examples, $m \leq n$, and thus we obtain $\mathcal{O}(n\cdot \log n)$ and $\mathcal{O}(n)$ upper bounds for expected-runtime
analysis, which are the asymptotically optimal bounds.
In all cases above, the worst-case analysis is completely ineffective as the worst-case bounds are infinite.
Moreover, consider Example~\ref{ex:channel}, where the optimal number of rounds is $n$ (i.e., one process every round,
which centralized Round-Robin schemes can achieve).
The randomized algorithm, with one shared variable, is a decentralized algorithm that achieves $O(n)$ expected
number of rounds (i.e., the optimal asymptotic expected-runtime complexity).


%% file: experiments.tex
\vspace{-1.5em}
\section{Experimental Results}\label{sect:experiments}
\vspace{-1em}

We consider the classical examples illustrated in Section~\ref{sec:motivatinguni}
and Section~\ref{sec:motivatingbi}.
In Table~\ref{tab:experiments} for experimental results we consider the following recurrence relations $G$:
{\sc R.-Sear.} corresponds to the recurrence relation (\ref{eq:relrandsearch}) for Example~\ref{ex:randsearch};
{\sc Q.-Sort} corresponds to the recurrence relation (\ref{eq:relquicksort}) for Example~\ref{ex:quicksort};
{\sc Q.-Select} corresponds to the recurrence relation (\ref{eq:relquickselect}) for Example~\ref{ex:quickselect};
{\sc Diam. A} (resp. {\sc Diam. B}) corresponds to the recurrence relation (\ref{eq:reldiametera}) (resp. the recurrence relation (\ref{eq:reldiameterb})) for Example~\ref{ex:diameter};
{\sc Sort-Sel.} corresponds to recurrence relation (\ref{eq:relsortselect}) for Example~\ref{ex:sortselect}, where we use the result from setting $\epsilon=0.01$ in {\sc Q.-Select};
{\sc Coupon} corresponds to the recurrence relation (\ref{eq:relcoupon}) for Example~\ref{ex:coupon};
{\sc Res. A} (resp. {\sc Res. B}) corresponds to the recurrence relation (\ref{eq:relresourcea}) (resp. the recurrence relation (\ref{eq:relresourceb})) for Example~\ref{ex:channel}.


In the table, $\mathfrak{f}$ specifies the input asymptotic bound, $\epsilon$ and Dec is the input which specifies either we use algorithm $\mbox{\sl UniDec}$ or the synthesis algorithm $\mbox{\sl UniSynth}$ with the given $\epsilon$ value, and $d$ gives the value synthesized w.r.t the given $\epsilon$ ($\checkmark$ for $\mbox{\sl yes}$).
We describe $d_{100}$ below.
We need approximation for constants such as $e$ and $\ln{2}$, and use the interval $[2.7182,2.7183]$
(resp., $[0.6931, 0.6932]$) for tight approximation of $e$ (resp., $\ln{2}$).

\smallskip\noindent{\em The value $d_{100}$.}
For our synthesis algorithm we obtain the value $d$. The optimal value of the associated constant with the asymptotic bound, denoted $d^*$, is defined as
follows.
For $z\geq 2$, let
$d_{z}:=\max\left\{\frac{T_G(n)-c}{\Sub(\mathfrak{f})(n)}\mid 2\le n\le z\right\}$
($c$ is from (\ref{eq:unirecurrel})).
Then the sequence $d_z$ is increasing in $z$, and its limit is the optimal constant, i.e., $d^* =\lim_{z \to \infty} d_z$.
We consider $d_{100}$ as a lower bound on $d^*$ to compare against the value of $d$ we synthesize.
In other words, $d_{100}$ is the minimal value such that (\ref{eq:uniguess}) holds for $1\le n\le 100$, whereas
for $d^*$ it must hold for all $n$, and hence $d^* \geq d_{100}$.
Our experimental results show that the $d$ values we synthesize for $\epsilon=0.01$ is quite close
to the optimal value.

We performed our experiments on Intel(R) Core(TM) i7-4510U CPU, 2.00GHz, 8GB RAM.
All numbers in Table~\ref{tab:experiments} are over-approximated up to $10^{-3}$, and the running time of all experiments are less than $0.02$ seconds.
From Table~\ref{tab:experiments}, we can see that optimal $d$ are effectively over-approximated.
For example, for {\sc Quick-Sort} (Eq.~(\ref{eq:relquicksort})) (i.e, {\sc Q.-Sort} in the table), our algorithm detects $d=4.051$ and the optimal one lies somewhere in $[3.172, 4.051]$.
The experimental results show that we obtain the results extremely efficiently (less than $1/50$-th of a second).
For further details see Table~\ref{tbl:detailedexperiments} in Appendix~\ref{app:experiments}.



\begin{table}[h]
\centering
\begin{tabular}{|c|c|c|c|c|c|c|c|c|c|}
\hline
Recur. Rel. &  $\mathfrak{f}$  & $\epsilon,\mbox{\sl Dec} $  &  $d$ & $d_{100}$   &
Recur. Rel. &  $\mathfrak{f}$  & $\epsilon,\mbox{\sl Dec} $  &  $d$ & $d_{100}$ \\
\hline
\multirow{5}{*}{{\sc R.-Sear.}}  & \multirow{5}{*}{$\ln{\mathfrak{n}}$} & $\mbox{\sl UniDec}$ & $\mbox{\sl \checkmark}$ & \multirow{5}{*}{$15.129$}
& \multirow{5}{*}{{\sc Sort-Sel.}}  & \multirow{5}{*}{$\mathfrak{n}\cdot\ln{\mathfrak{n}}$} & $\mbox{\sl UniDec}$ &  $\mbox{\sl \checkmark}$ & \multirow{5}{*}{$16.000$} \\
\cline{3-4}\cline{8-9} & &  $0.5$  & $40.107$ &  &  & &  $0.5$  &      $50.052$  &  \\
\cline{3-4}\cline{8-9} & &  $0.3$  & $28.363$ &  &  & &  $0.3$  &      $24.852$  &  \\
\cline{3-4}\cline{8-9} & &  $0.1$  & $21.838$ &  &  & &  $0.1$  &      $17.313$  &  \\
\cline{3-4}\cline{8-9} & &  $0.01$ & $19.762$ &  &  & &  $0.01$ &      $16.000$  &  \\
\hline
\multirow{5}{*}{{\sc Q.-Sort}}  & \multirow{5}{*}{$\mathfrak{n}\cdot\ln{\mathfrak{n}}$} & $\mbox{\sl UniDec}$  & $\mbox{\sl \checkmark}$ & \multirow{5}{*}{$3.172$} & \multirow{5}{*}{{\sc Coupon}}  & \multirow{5}{*}{$\mathfrak{n}\cdot\ln{\mathfrak{m}}$} & $\mbox{\sl UniDec}$ & $\mbox{\sl \checkmark}$  & \multirow{5}{*}{$0.910$}   \\
\cline{3-4}\cline{8-9} & & $0.5$   & $9.001$ & &  & & $0.5$  &     $3.001$  &  \\
\cline{3-4}\cline{8-9} & & $0.3$   & $6.143$ & &  & & $0.3$  &     $1.858$  &  \\
\cline{3-4}\cline{8-9} & & $0.1$   & $4.556$ & &  & & $0.1$  &     $1.223$  &  \\
\cline{3-4}\cline{8-9} & & $0.01$  & $4.051$ & &  & & $0.01$ &     $1.021$  &  \\
\hline
\multirow{5}{*}{{\sc Q.-Select}}  & \multirow{5}{*}{$\mathfrak{n}$}
 & $\mbox{\sl UniDec}$  &  $\mbox{\sl \checkmark}$  & \multirow{5}{*}{$7.909$} & \multirow{5}{*}{{\sc Res. A}} & \multirow{5}{*}{$\mathfrak{n}\cdot\ln{\mathfrak{m}}$}   & $\mbox{\sl UniDec}$  & $\mbox{\sl \checkmark}$   & \multirow{5}{*}{$2.472$}  \\
\cline{3-4}\cline{8-9} & &  $0.5$  & $17.001$  & & & &  $0.5$ &  $6.437$  &   \\
\cline{3-4}\cline{8-9} & &  $0.3$  & $11.851$  & & & & $0.3$  &  $4.312$  & \\
\cline{3-4}\cline{8-9} & &  $0.1$  &  $9.001$  & & & & $0.1$  &  $3.132$  &\\
\cline{3-4}\cline{8-9} & &  $0.01$ &  $8.091$  & & & & $0.01$ &  $2.756$  &\\
\hline
\multirow{5}{*}{{\sc Diam. A}} & \multirow{5}{*}{$\mathfrak{n}\cdot\ln{\mathfrak{n}}$}
& $\mbox{\sl UniDec}$  & $\mbox{\sl \checkmark}$  &   \multirow{5}{*}{$4.525$} &
\multirow{5}{*}{{\sc Res. B}}  & \multirow{5}{*}{$\mathfrak{m}$}
& $\mbox{\sl UniDec}$ &  $\mbox{\sl \checkmark}$ & \multirow{5}{*}{$2.691$}   \\
\cline{3-4}\cline{8-9} & &  $0.5$  & $9.001$ & &  & &  $0.5$  & $6.437$  & \\
\cline{3-4}\cline{8-9} & &  $0.3$  & $6.143$ & &  & &  $0.3$  & $4.312$  &  \\
\cline{3-4}\cline{8-9} & &  $0.1$  & $4.556$ & &  & &  $0.1$  & $3.132$  &  \\
\cline{3-4}\cline{8-9} & &  $0.01$ & $4.525$ & &  & &  $0.01$ & $2.756$  & \\
\hline
\multirow{5}{*}{{\sc Diam. B}}  & \multirow{5}{*}{$\mathfrak{n}$} & $\mbox{\sl UniDec}$ & $\mbox{\sl \checkmark}$ & \multirow{5}{*}{$5.918$} &
\multirow{5}{*}{{-}}  & \multirow{5}{*}{-} & - & - & \multirow{5}{*}{-}  \\
\cline{3-4}\cline{8-9} &  & $0.5$  &  $13.001$ & &  &  & - & - &\\
\cline{3-4}\cline{8-9} &  & $0.3$  &  $9.001$  & &  &  & - & - &\\
\cline{3-4}\cline{8-9} &  & $0.1$  &  $6.778$  & &  &  & - & - &\\
\cline{3-4}\cline{8-9} &  & $0.01$ &  $6.071$  & &  &  & - & - & \\
\hline
\end{tabular}
\caption{Experimental results where all running times (averaged over $5$ runs) are less than $0.02$ seconds, between
$0.01$ and $0.02$ in all cases.}
\label{tab:experiments}
\end{table}


%% file: conclu.tex
\vspace{-1.5em}
\section{Related Work}
\vspace{-1em}

Automated program analysis is a very important problem with a long tradition~\cite{DBLP:journals/cacm/Wegbreit75}.
The following works consider various approaches for automated worst-case
bounds~\cite{DBLP:conf/aplas/HoffmannH10,DBLP:conf/esop/HoffmannH10,DBLP:conf/popl/HofmannJ03,DBLP:conf/esop/HofmannJ06,DBLP:conf/csl/HofmannR09,DBLP:conf/fm/JostLHSH09,DBLP:conf/popl/JostHLH10,Hoffman1,DBLP:conf/icfp/AvanziniLM15,DBLP:conf/se/SinnZV16} for amortized analysis,
and the SPEED project~\cite{SPEED1,SPEED2,DBLP:conf/cav/GulavaniG08}
for non-linear bounds using abstract interpretation.
All these works focus on the worst-case analysis, and do not consider
expected-runtime analysis.

Our main contribution is automated analysis of recurrence relations.
Approaches for recurrence relations have also been considered in the
literature.
Wegbreit~\cite{DBLP:journals/cacm/Wegbreit75} considered solving recurrence relations through either simple difference equations or generating functions.
Zimmermann and Zimmermann~\cite{Zimmermann1989} considered solving recurrence relations by transforming them into difference equations.
Grobauer~\cite{DBLP:conf/icfp/Grobauer01} considered generating recurrence relations from DML
for the worst-case analysis.
Flajolet~\emph{et al.}~\cite{DBLP:journals/dam/FlajoletGT92} considered allocation problems.
Flajolet~\emph{et al.}~\cite{DBLP:journals/tcs/FlajoletSZ91} considered solving recurrence relations
for randomization of combinatorial structures (such as trees) through generating functions.
The COSTA project~\cite{DBLP:journals/entcs/AlbertAGGPRRZ09,DBLP:conf/sas/AlbertAGP08,DBLP:conf/esop/AlbertAGPZ07}
transforms Java bytecode into recurrence relations and solves them through ranking functions.
Moreover, The PURRS tool~\cite{BagnaraPZZ05} addresses finite linear recurrences (with bounded summation), and some restricted linear infinite recurrence relations (with unbounded summation).
Our approach is quite different because we consider analyzing recurrence relations
arising from randomized algorithms and expected-runtime analysis through over-approximation of unbounded summations through integrals, whereas
previous approaches either consider recurrence relations for worst-case bounds or combinatorial structures, or use generating functions or difference equations to solve the recurrence relations.

For intraprocedural analysis ranking functions have been widely studied~\cite{BG05,DBLP:conf/cav/BradleyMS05,DBLP:conf/tacas/ColonS01,DBLP:conf/vmcai/PodelskiR04,DBLP:conf/pods/SohnG91,DBLP:conf/vmcai/Cousot05,DBLP:journals/fcsc/YangZZX10,DBLP:journals/jossac/ShenWYZ13},
which have then been extended to non-recursive probabilistic programs as
ranking supermartingales~\cite{SriramCAV,HolgerPOPL,DBLP:conf/popl/ChatterjeeFNH16,DBLP:conf/cav/ChatterjeeFG16,ChatterjeeNZ2017,DBLP:journals/corr/ChatterjeeF17}.
Such approaches are related to almost-sure termination, and not deriving
optimal asymptotic expected-runtime bounds (such as $\mathcal{O}(\log n)$, $\mathcal{O}(n \log n)$).

Proof rules have also been considered for recursive (probabilistic)
programs in~\cite{DBLP:journals/fac/Hesselink94,JonesPhdThesis,DBLP:conf/lics/OlmedoKKM16},
but these methods cannot be automated and require manual proofs.

\vspace{-1.5em}
\section{Conclusion}
\vspace{-1em}
In this work we considered efficient algorithms for automated analysis of randomized
recurrences for logarithmic, linear, and almost-linear bounds.
Our work gives rise to a number of interesting questions.
First, an interesting theoretical direction of future work would be to consider
more general randomized recurrence relations (such as with more than two variables,
or interaction between the variables).
While the above problem is of theoretical interest, most interesting examples are
already captured in our class of randomized recurrence relations as mentioned above.
Another interesting practical direction would be automated techniques
to derive recurrence relations from randomized recursive programs.

\vspace{-1.5em}
\subsubsection*{Acknowledgements}
We thank all reviewers for valuable comments.
The research is partially supported by Vienna Science and Technology Fund (WWTF) ICT15-003,
Austrian Science Fund (FWF) NFN Grant No. S11407-N23 (RiSE/SHiNE), ERC Start grant (279307: Graph Games), the Natural Science Foundation of China (NSFC) under Grant No. 61532019 and the CDZ project CAP (GZ 1023).

%% file: appendix.tex
\appendix

\section{Omitted Details for Section~\ref{sec:motivatinguni}}\label{sect:recurreldetails}

\lstset{language=prog}
\lstset{tabsize=3}
\newsavebox{\progsearch}
\begin{lrbox}{\progsearch}
\begin{lstlisting}[mathescape]
$\mathsf{randsearch}(ar,i,j,d)$ {
1:  if ($i=j$ and $ar[i]\ne d$)
2:   return $-1$;
3:  else if ($i=j$ and $ar[i]=d$)
4:   return $i$;
5:  else
6:   $k\leftarrow \mathrm{uniform}(i,j)$;

7:   if ($ar[k]=d$)
8:    return $k$;
9:   else if ($ar[k]<d$ and $k<j$)
10:   return $\mathsf{randsearch}(ar, k+1, j,d)$;
11:  else if ($ar[k]>d$ and $i<k$)
12:   return $\mathsf{randsearch}(ar,1,k-1,d)$;
13:  else
14:   return $-1$;
     end if
    end if
}
\end{lstlisting}
\end{lrbox}

\begin{figure}
\centering
\usebox{\progsearch}
\caption{Sherwood's {\sc Randomized-Search}}
\label{fig:randsearch}
\end{figure}

\noindent{\bf Example~\ref{ex:randsearch}.} [{\sc Randomized-Search}]
Consider the Sherwood's {\sc Randomized-Search\ } algorithm (cf.~\cite[Chapter~9]{McConnellbook}) depicted in Fig.~\ref{fig:randsearch}.
The algorithm checks whether an integer value $d$ is present within the index range $[i,j]$ ($0\le i\le j$)
in an integer array $ar$ which is sorted in increasing order and is without duplicate entries.
The algorithm outputs either the index for $d$ in $ar$ or $-1$ meaning that $d$ is not present in
the index range $[i,j]$ of  $ar$.

The description of the pseudo-code is as follows.
The first four lines deal with the base case when there is only one index in the index range.
The remaining lines deal with the recursive case:
in line 6, an index $k$ is uniformly sampled from $\{i,i+1,\dots,j\}$;
line 7--8 check whether $k$ is the output;
line 9--12 perform the recursive calls depending on whether $ar[k]<d$ or not;
finally, line 13--14 handle the case when $d<ar[i]$ or $d>ar[j]$.

Let $T:\Nset\rightarrow\Nset$ be the function such that for any $n\in\Nset$, we have
$T(n)$ is the supremum of the expected execution times upon all inputs $(ar,i,j)$ with $j-i+1=n$.
We derive a recurrence relation for $T$ as follows.
Let $n\in\Nset$ and $(ar,i,j), d$ be any input such that $n=j-i+1$.
We clarify two cases below:
\begin{enumerate}
\item there exists an $i\le k^*< j$ such that $ar[k^*]\le d < ar[k^*+1]$, where $ar[j+1]$ is interpreted $\infty$ here;
\item $ar[j]\le d$ or $d< ar[i]$.
\end{enumerate}
In both cases, we have $T(1)=1$.
In Case 1, we deduce from the pseudo-code in Fig.~\ref{fig:randsearch} that
\begin{displaymath}
T(n)\le 6+\frac{1}{n}\cdot \max\limits_{1\le \ell^*< n} \left(\displaystyle\sum_{\ell=1}^{\ell^*} T(n-\ell)+ \displaystyle\sum_{\ell=\ell^*+1}^{n} T(\ell-1)\right)
\end{displaymath}
for all $n\ge 2$, where the maximum ranges over all $\ell^*:=k^*-i+1$'s.
In Case 2, similarly we deduce that
\begin{displaymath}
T(n)\le 6+\frac{1}{n}\cdot\max\left\{\displaystyle\sum_{\ell=1}^{n-1} T(n-\ell), ~~\displaystyle\sum_{\ell=2}^{n} T(\ell-1)\right\}
\end{displaymath}
Thus a preliminary version $G'$ of the recurrence relation is $\mathrm{T}(1)=1$ and
\[
\mathrm{T}(n)=6+\frac{1}{n}\cdot\max\limits_{1\le \ell^*< n} \left(\displaystyle\sum_{\ell=1}^{\ell^*} \mathrm{T}(n-\ell)+ \displaystyle\sum_{\ell=\ell^*+1}^{n} \mathrm{T}(\ell-1)\right) \\
\]
for all $n\ge 2$.
Let $T':\Nset\rightarrow\Rset$ be the unique solution to $G'$.
Then from the fact that $T'(2)\ge T'(1)$, by induction $T'$ is monotonically increasing.
Thus the maximum
\[
\max\limits_{1\le \ell^*<n} \left(\displaystyle\sum_{\ell=1}^{\ell^*} T'(n-\ell)+ \displaystyle\sum_{\ell=\ell^*+1}^{n} T'(\ell-1)\right)
\]
is attained at $\ell^*=\left\lfloor\frac{n}{2}\right\rfloor$ for all $n\ge 2$.
Then $G'$ is transformed into our final recurrence relation  as follows:
\[
\begin{cases}
\mathrm{T}(\mathfrak{n})=6+\frac{1}{\mathfrak{n}}\cdot\left( \displaystyle\sum_{\mathfrak{j}=\left\lceil\frac{\mathfrak{n}}{2}\right\rceil}^{\mathfrak{n}-1}\mathrm{T}(\mathfrak{j})+ \displaystyle\sum_{\mathfrak{j}=\left\lfloor\frac{\mathfrak{n}}{2}\right\rfloor}^{\mathfrak{n}-1} \mathrm{T}(\mathfrak{j})\right) \\
\mathrm{T}(1)=1
\end{cases}.
\]
We note that the worst-case complexity for this algorithm is $\Theta(n)$.\qed

\lstset{language=prog}
\lstset{tabsize=3}
\newsavebox{\progsort}
\begin{lrbox}{\progsort}
\begin{lstlisting}[mathescape]
$\mathsf{quicksort}(ar,i,j)$ {
1: if ($i<j$)
2:   $k\leftarrow \mathrm{uniform}(i,j)$;
3:   $m\leftarrow \mathsf{pivot}(ar,i,j,ar[k])$;

4:   if ($i\le m-1$)
5:     $\mathsf{quicksort}(ar,i,m-1)$;
     end if

6:   if ($m+1\le j$)
7:     $\mathsf{quicksort}(ar,m+1,j)$;
     end if
   end if
}
\end{lstlisting}
\end{lrbox}

\begin{figure}
\centering
\usebox{\progsort}
\caption{Randomized {\sc Quick-Sort}}
\label{fig:quicksort}
\end{figure}

\noindent{\em Example \ref{ex:quicksort}.}[{\sc Quick-Sort}]
Consider the {\sc Quick-Sort} algorithm~\cite[Chapter~7]{DBLP:books/daglib/0023376} depicted in Fig.~\ref{fig:quicksort}, where every input $(ar,i,j)$ is assumed to satisfy that $0\le i\le j$ and $ar$ is an array of integers which does not contain duplicate numbers.

The description of the pseudo-code is as follows:
first, line 2 samples an integer uniformly from $\{i,\dots, j\}$;
then, line 3 calls a subroutine $\mathsf{pivot}$ which (i) rearranges $ar$ such that integers in $ar$ which are less than $ar[k]$ come first,
then $ar[k]$, and finally integers in $ar$ greater than $ar[k]$, and (ii) outputs the new index $m$ of $ar[k]$ in $ar$; and
finally, lines 4--7 handle recursive calls to sub-arrays.

From the pseudo-code, the following recurrence relation is easily obtained:
\[
\mathrm{T}(\mathfrak{n})=2\cdot\mathfrak{n}+ 2\cdot (\sum_{\mathfrak{j}=1}^{\mathfrak{n}-1} \mathrm{T}(\mathfrak{j}))/{\mathfrak{n}}
\]
where $\mathrm{T}(\mathfrak{n})$ represents the maximal expected execution time where
$\mathfrak{n}$ is the array length and the execution time of {\em pivoting} is represented
by $2\cdot \mathfrak{n}$.
We note that the worst-case complexity for this algorithm is $\Theta(n^2)$.\qed

\lstset{language=prog}
\lstset{tabsize=3}
\newsavebox{\progdiameter}
\begin{lrbox}{\progdiameter}
\begin{lstlisting}[mathescape]
$\mathsf{diameter}(S)$ {
1: if ($|S|=1$)
2:  return 0;
   else
3:  $p\leftarrow \mathrm{uniform}(S)$;
4:  $d\leftarrow \max_{p'\in S} \mbox{\sl dist}(p,p')$
5:  $U\leftarrow \bigcap_{p'\in S}\{p''\in\Rset^3 \mid  \mbox{\sl dist}(p'',p')\le d\}$
6:  $S'\leftarrow S\backslash U$

7:  if ($S'=\emptyset$)
8:   return $d$
9:   else
10:   return $\mathsf{diameter}(S')$
     end if
    end if
}
\end{lstlisting}
\end{lrbox}

\lstset{language=prog}
\lstset{tabsize=3}
\newsavebox{\progselect}
\begin{lrbox}{\progselect}
\begin{lstlisting}[mathescape]
$\mathsf{quickselect}(ar,i,j,d)$ {
1: if ($i=j$) return $a[i]$;
2: else
3:  $k\leftarrow \mathrm{uniform}(i,j)$;
4:  $m\leftarrow \mathsf{pivot}(ar,i,j,ar[k])$;

5:  if ($m-i+1=d$)
6:   return $ar[m]$;
7:  else if ($m-i+1<d$)
8:   return $\mathsf{quickselect}(ar,m+1,j,d)$;
9:  else if ($m-i+1>d$)
10:  return $\mathsf{quickselect}(ar,i,m-1,d)$;
    end if
   end if
}
\end{lstlisting}
\end{lrbox}
\begin{figure}
\begin{minipage}{0.55\textwidth}
\centering
\usebox{\progselect}
\caption{Randomized {\sc Quick-Select}}
\label{fig:quickselect}
\end{minipage}
\begin{minipage}{0.45\textwidth}
\centering
\usebox{\progdiameter}
\caption{{\sc Diameter-Computation}}
\label{fig:diameter}
\end{minipage}
\end{figure}

\noindent{\bf Example~\ref{ex:quickselect}.} [{\sc Quick-Select}]
Consider the {\sc Quick-Select} algorithm (cf.~\cite[Chapter~9]{DBLP:books/daglib/0023376}) depicted in Fig.~\ref{fig:quickselect} which upon any input $(ar,i,j)$ and $d$ such that
$0\le i\le j$, $1\le d\le j-i+1$ and $ar$ contains no duplicate integers, finds the $d$-th largest integer in $ar$.
Note that for an array of size $n$, and $d=n/2$, we have the {\sc Median-Find} algorithm.

The description of the pseudo-code is as follows: line 1 handles the base case;
line 3 starts the recursive case by sampling $k$ uniformly from $\{i,\dots,j\}$;
line 4 rearranges $ar$ and returns an $m$ in the same way as $\mathsf{pivot}$ in {\sc Quick-Sort}
(cf. Example~\ref{ex:quicksort});
line 5 handles the case when $ar[k]$ happens to be the $d$-th largest integer in $ar$;
and finally, line 7--10 handle the recursive calls.

Let $T:\Nset\rightarrow\Nset$ be the function such that for any $n\in\Nset$, we have $T(n)$ is the supremum of the
expected execution times upon all inputs $(ar,i,j)$ with $j-i+1=n$.
By an analysis on where the $d$-th largest integer lies in $ar$ which is similar to the analysis on $d$ in Example~\ref{ex:randsearch},
a preliminary recurrence relation is obtained such that $\mathrm{T}(1)=1$ and
\[
\mathrm{T}(n)=4+2\cdot n+\frac{1}{n}\cdot\max\limits_{1\le \ell^*\le n} \left(\displaystyle\sum_{\ell=1}^{\ell^*-1} \mathrm{T}(n-\ell)+ \displaystyle\sum_{\ell=\ell^*+1}^{n} \mathrm{T}(\ell-1)\right). \\
\]
By similar monotone argument in Example~\ref{ex:randsearch}, the maximum of the right-hand-side expression above is attained at $\ell^*=\left\lfloor\frac{n+1}{2}\right\rfloor$ for all $n\ge 2$.
By the fact that $\left\lfloor\frac{n+1}{2}\right\rfloor=\left\lceil \frac{n}{2}\right\rceil$ for all $n\ge 2$, the following recurrence relation is obtained:
\[
\begin{cases}
\mathrm{T}(\mathfrak{n})=4+2\cdot\mathfrak{n}+
\frac{1}{\mathfrak{n}}\cdot \left(\displaystyle\sum_{\mathfrak{j}=\left\lfloor \frac{n}{2}\right\rfloor+1}^{n-1} \mathrm{T}(\mathfrak{j})+ \displaystyle\sum_{\mathfrak{j}=\left\lceil \frac{n}{2}\right\rceil}^{\mathfrak{n}-1} \mathrm{T}(\mathfrak{j})\right)\\
\mathrm{T}(1)=1
\end{cases}
\]
To fit our univariate recurrence expression, we use over-approximation, and the final recurrence relation for this example is
\[
\begin{cases}
\mathrm{T}(\mathfrak{n})\!=\!4+2\cdot\mathfrak{n}+
\frac{1}{\mathfrak{n}}\cdot \left(\displaystyle\sum_{\mathfrak{j}=\left\lfloor \frac{n}{2}\right\rfloor}^{n-1} \mathrm{T}(\mathfrak{j})+ \displaystyle\sum_{\mathfrak{j}=\left\lceil \frac{n}{2}\right\rceil}^{\mathfrak{n}-1} \mathrm{T}(\mathfrak{j})\right)\\
\mathrm{T}(1)=1
\end{cases}.
\]
We note that the worst-case complexity for this algorithm is $\Theta(n^2)$.\qed

\noindent{\em Example~\ref{ex:diameter}.}[{\sc Diameter-Computation}]
Consider the {\sc Diameter-Computation} algorithm (cf.~\cite[Chapter 9]{DBLP:books/cu/MotwaniR95}) to compute the diameter of an input finite set $S$ of three-dimensional points.
A pseudo-code to implement this is depicted in Fig.~\ref{fig:diameter}.
The description of the pseudo-code is as follows:
line 1--2 handle the base case;
line 3 samples a point $p$ uniformly from $S$;
line 4 calculates the maximum distance in $S$ from $p$;
line 5 calculates the intersection of all balls centered at points in $S$ with uniform radius $d$;
line 6 calculates the set of points outside $U$;
lines 7--8 handle the situation $S'=\emptyset$ which implies that $d$ is the diameter;
lines 9--10 handle the recursive call to $S'$.
Due to uniform choice of $p$ at line 3, the size of $S'$ is uniformly in $[0,|S|-1]$;
it then follows a pivoting (similar to that in Example~\ref{ex:quickselect} and Example~\ref{ex:quicksort}) by line $5$
w.r.t the linear order over $\{\max_{p'\in S}{\mbox{\sl dist}(p,p')}\mid p\in S\}$.
Lines 5--6 can be done in $\mathcal{O}(|S|\cdot \log{|S|})$ time for Euclidean distance,
and  $\mathcal{O}(|S|)$ time for $L_1$ metric~\cite{DBLP:books/cu/MotwaniR95}.

Depending on Eucledian or $L_1$ metric we obtain two different recurrence relations.
For Eucledian we have the following relation:
\[
\mathrm{T}(\mathfrak{n})=2+\mathfrak{n}+ 2\cdot \mathfrak{n}\cdot\ln{\mathfrak{n}} + (\sum_{\mathfrak{j}=1}^{\mathfrak{n}-1} \mathrm{T}(\mathfrak{j}))/{\mathfrak{n}} ;
\]
with the execution time for  lines 5--6 being taken to be $2\cdot \mathfrak{n}\cdot\ln{\mathfrak{n}}$, and
and for $L_1$ metric we have the following relation:
\[
\mathrm{T}(\mathfrak{n})=2+\mathfrak{n}+ 2\cdot \mathfrak{n} + (\sum_{\mathfrak{j}=1}^{\mathfrak{n}-1} \mathrm{T}(\mathfrak{j}))/{\mathfrak{n}} \\
\]
with the execution time for lines 5--6 being taken to be
$2\cdot \mathfrak{n}$.
We note that the worst-case complexity for this algorithm is as follows:
for Euclidean metric it is $\Theta(n^2 \cdot \log n)$ and for the $L_1$ metric
it is $\Theta(n^2)$.\qed

\lstset{language=prog}
\lstset{tabsize=3}
\newsavebox{\progsortselect}
\begin{lrbox}{\progsortselect}
\begin{lstlisting}[mathescape]

$\mathsf{sortbyselect}(ar,i,j)$ {
1:  if ($i<j$)
2:    $m\leftarrow \mathsf{quickselect}(ar,i,j,\lfloor\frac{j-i+1}{2}\rfloor)$;
3:    if ($i< m-1$)
4:      $\mathsf{sortbyselect}(ar,i,m-1)$;
      end if

5:    if ($m+1<j$)
6:      $\mathsf{sortbyselect}(ar,m+1,j)$;
      end if
    end if
}
\end{lstlisting}
\end{lrbox}
\begin{figure}
\centering
\usebox{\progsortselect}
\caption{Sorting with {\sc Quick-Select}}
\label{fig:sortselect}
\end{figure}

\noindent{\em Example \ref{ex:sortselect}.}[Sorting with {\sc Quick-Select}]
Consider a sorting algorithm depicted in Fig.~\ref{fig:sortselect}
which selects the median through the {\sc Quick-Select} algorithm.
The recurrence relation is directly obtained as follows:
\[
\mathrm{T}(\mathfrak{n})=4+ T^*(\mathfrak{n})+\mathrm{T}\left(\lfloor{\mathfrak{n}}/{2}\rfloor\right)+\mathrm{T}\left(\lceil{\mathfrak{n}}/{2}\rceil\right)
\]
where $T^*(\centerdot)$ is an upper bound on the expected running time of {\sc Quick-Select}
(cf. Example~\ref{ex:quickselect}).
We note that the worst-case complexity for this algorithm is $\Theta(n^2)$.\qed

\section{Omitted Details for Section~\ref{sec:motivatingbi}}\label{app:motivatingbi}

\noindent{\bf Example~\ref{ex:coupon}.}[{\sc Coupon-Collector}]
Consider the {\sc Coupon-Collector} problem~\cite[Chapter~3]{DBLP:books/cu/MotwaniR95} with $n$ different types of coupons ($n\in\Nset$).
The randomized process proceeds in rounds: at each round, a coupon is collected uniformly at random from the coupon types
(i.e., each coupon type is collected with probability $\frac{1}{n}$); and
the rounds continue until all the $n$ types of coupons are collected.
We model the rounds as a recurrence relation with two variables
$\mathfrak{n},\mathfrak{m}$, where $\mathfrak{n}$ represents the total number
of coupon types and $\mathfrak{m}$ represents the remaining number of uncollected
coupon types.
The recurrence relation is as follows:
\[
\mathrm{T}(\mathfrak{n},1)=\mathfrak{n}\cdot 1; \qquad
\mathrm{T}(\mathfrak{n},\mathfrak{m})=\mathfrak{n}/{\mathfrak{m}}+ \mathrm{T}(\mathfrak{n},\mathfrak{m}-1)
\]
where $\mathrm{T}(\mathfrak{n},\mathfrak{m})$ is the expected number of rounds, $\frac{\mathfrak{n}}{\mathfrak{m}}$
represents the expected number of rounds to collect a new (i.e., not-yet-collected) coupon type
when there are still $\mathfrak{m}$ type of coupons to be collected, and $\mathfrak{n}$ (for $\mathrm{T}(\mathfrak{n},1)$)
represents the expected number of rounds to collect a new coupon type when there is only one new coupon type to be collected.
We note that the worst-case complexity for this process is $\infty$.\qed

\noindent{\bf Example~\ref{ex:channel}.}[{\sc Channel-Conflict Resolution}]
We consider two network scenarios in which $n$ clients are trying to get
access to a network channel.
This problem is also called the {\sc Resource-Contention Resolution}~\cite[Chapter~13]{Kleinbergbook}.
In this problem, if more than one client tries to access the channel,
then no client can access it, and if exactly one client requests access to
the channel, then the request is granted.
While centralized deterministic algorithms exist (such as Round-Robin) for
the problem, to be implemented in a distributed or concurrent setting,
randomized algorithms are necessary.

\smallskip\noindent{\em Distributed setting.}
In the distributed setting, the clients do not share any information.
In this scenario, in each round, every client requests an access to the
channel with probability $\frac{1}{n}$.
We are interested in the expected number of rounds until every client gets
at least one access to the channel.
At each round, let $m$ be the number of clients who have not got any access.
Then the probability that a new client (from the $m$ clients) gets the access is $m\cdot \frac{1}{n}\cdot (1-\frac{1}{n})^{n-1}$.
Thus, the expected rounds that a new client gets the access is $\frac{n}{m}\cdot \frac{1}{(1-\frac{1}{n})^{n-1}}$.
Since the sequence $\left\{(1-\frac{1}{n})^{n-1}\right\}_{n\in\Nset}$ converges decreasingly to $\frac{1}{e}$ when
$n\rightarrow\infty$, this expected time is no greater than
$e\cdot\frac{n}{m}$.
Then for this scenario, we obtain an over-approximating recurrence relation
\[
\mathrm{T}(\mathfrak{n},1)=\mathfrak{n}\cdot 1; \qquad
\mathrm{T}(\mathfrak{n},\mathfrak{m})=(\mathfrak{n}\cdot{e})/{\mathfrak{m}}+ \mathrm{T}(\mathfrak{n},\mathfrak{m}-1)
\]
for the expected rounds until which every client gets at least one access to the channel.
Note that in this setting no client has any information about any other client.

\smallskip\noindent{\em Concurrent setting.}
In the concurrent setting, the clients share one variable, which is the number
of clients which has not yet been granted access.
Also in this scenario, once a client gets an access the client does not
request for access again.
Moreover, the shared variable represents the number of clients $m$
that have not yet got access.
In this case, in reach round a client that has not access to the channel yet,
requests access to the channel with probability $\frac{1}{m}$.
Then the probability that a new client gets the access becomes
$m\cdot \frac{1}{m}\cdot (1-\frac{1}{m})^{m-1}$.
It follows that the expected time that a new client gets the access becomes
$\frac{1}{(1-\frac{1}{m})^{m-1}}$
which is smaller than $e$.
Then for this scenario, we obtain an over-approximating recurrence relation
\[
\mathrm{T}(\mathfrak{n},1)=1\cdot 1; \qquad
\mathrm{T}(\mathfrak{n},\mathfrak{m})=1\cdot e+ \mathrm{T}(\mathfrak{n},\mathfrak{m}-1)
\]
We also note that the worst-case complexity for both is $\infty$.\qed

\section{Details for Overapproximations}\label{app:overapprox}

To prove results for overapproximations for recurrence expressions, we need the following well-known theorem.

\begin{theorem}[Taylor's Theorem (with Lagrange's Remainder){~\cite[Chapter 6]{BasicCalculus}}]
For any function $f:[a,b]\rightarrow \Rset$ ($a,b\in\Rset$ and $a<b$), if $f$ is ($k+1$)-order differentiable, then for all $x\in [a,b]$, there exists a $\xi\in (a,x)$ such that
\[
f(x)=\left(\sum_{j=0}^k \frac{f^{(j)}(a)}{j!}\cdot (x-a)^j\right)+\frac{f^{(k+1)}(\xi)}{(k+1)!}\cdot (x-a)^{k+1}~~.
\]
\end{theorem}

We also recall that
\[
\sum_{j=1}^{\infty}\frac{1}{j^2}=\frac{\pi^2}{6}\mbox{ and }\sum_{j=1}^{\infty}\frac{1}{j^3}=\alpha
\]
where $\alpha$ is the Ap\'{e}ry's constant which lies in $[1.2020,1.2021]$.

Moreover, we have the following result using integral-by-part technique
and Newton-Leibniz Formula.

\begin{lemma}\label{lemm:basiccalculus}
For all $a,b\in (0,\infty)$ such that $a<b$, the following assertions hold:
\begin{eqnarray*}
\displaystyle
(1)\ \int_a^b \frac{1}{x}\,\mathrm{d}x  =   \ln{x}{\Big|}_{a}^b \enspace ; \displaystyle
~~~(2)\  \int_a^b \ln{x}\,\mathrm{d}x  =  \left(x\cdot\ln{x}-x\right){\Big|}_{a}^b \enspace\\[1ex]
\displaystyle
(3)\ \int_a^b x\cdot\ln{x}\,\mathrm{d}x  =  \left(\frac{1}{2}\cdot x^2\cdot\ln{x}-\frac{1}{4}\cdot x^2\right){\Big|}_{a}^b~~ \enspace.
\end{eqnarray*}
\end{lemma}

Furthermore, we need the following simple lemmas.
The following lemma provides a tight approximation for floored expressions, the proof of which is a simple case distinction between even and odd cases.

\begin{lemma}\label{lemm:flooroverapprox}
For all natural numbers $n$, we have
$\frac{n-1}{2}\le \left\lfloor\frac{n}{2}\right\rfloor\le \frac{n}{2}\le \left\lceil\frac{n}{2}\right\rceil \le \frac{n+1}{2}$~~.
\end{lemma}

The following lemma handles over-approximation of simple summations.

\begin{lemma}\label{lemm:sumoverapprox}
For any natural number $n\ge 2$ and real number $c$, one has that
$\frac{\sum_{j=1}^{n-1} c}{n}\le c\mbox{ and }\frac{\left( \sum_{j=\left\lceil\frac{\mathfrak{n}}{2}\right\rceil}^{n-1} c+ \sum_{j=\left\lfloor\frac{\mathfrak{n}}{2}\right\rfloor}^{n-1} c\right)}{n}\le c$~~.
\end{lemma}

Then we prove the following two propositions.

\textbf{Proposition~\ref{prop:lnflooroverapprox}.}
For any natural number $n\ge 2$, we have
\[
(1)\ \ln{n}-\ln{2}-\frac{1}{n-1}\le \ln{\left\lfloor \frac{n}{2}\right\rfloor}\le \ln{n}-\ln{2} \enspace ;
\]
\[
(2)\ \ln{n}-\ln{2}\le \ln{\left\lceil \frac{n}{2}\right\rceil}\le \ln{n}-\ln{2}+\frac{1}{n}~~.
\]
\begin{proof}
Let $n\ge 2$ be a natural number. The first argument comes from the facts that
\[
\ln{\left\lfloor \frac{n}{2}\right\rfloor}\le \ln{\frac{n}{2}}=\ln{n}-\ln{2}
\]
and
\begin{align*}
\ln{\left\lfloor \frac{n}{2}\right\rfloor} & \ge \ln{\frac{n-1}{2}} \\
& = \ln{\frac{n}{2}}-\left(\ln{\frac{n}{2}}-\ln{\frac{n-1}{2}}\right) \\
& = \ln{n}-\ln{2}-\frac{1}{2}\cdot \frac{1}{\xi_n}~~\left(\xi_n\in\left(\frac{n-1}{2},\frac{n}{2}\right)\right)\\
& \ge \ln{n}-\ln{2}-\frac{1}{n-1}
\end{align*}
where we use the fact that
\[
\left\lfloor \frac{n}{2}\right\rfloor\ge \frac{n-1}{2}
\]
and $\xi_n$ is obtained from Taylor's Theorem.
The second argument comes from the facts that
\[
\ln{\left\lceil \frac{n}{2}\right\rceil}\ge \ln{\frac{n}{2}}=\ln{n}-\ln{2}
\]
and
\begin{align*}
\ln{\left\lceil \frac{n}{2}\right\rceil} & \le \ln{\frac{n+1}{2}}\\
& =\ln{\frac{n}{2}}+\left(\ln{\frac{n+1}{2}}-\ln{\frac{n}{2}} \right)\\
& =\ln{n} -\ln{2} + \left(\ln{\frac{n+1}{2}}-\ln{\frac{n}{2}} \right)\\
& =\ln{n}-\ln{2}+\frac{1}{2}\cdot \frac{1}{\xi'_n}\quad \left(\xi'_n\in \left(\frac{n}{2}, \frac{n+1}{2}\right)\right)\\
& \le \ln{n}-\ln{2}+\frac{1}{n}
\end{align*}
where the first inequality is due to the fact that
\[
\left\lceil \frac{n}{2}\right\rceil\le \frac{n+1}{2}
\]
and $\xi'_n$ is obtained from Taylor's Theorem.\qed
\end{proof}

\textbf{Proposition~\ref{prop:nminusoneoverapprox}.}
For any natural number $n\ge 2$, we have
\[
\ln{n}-\frac{1}{n-1}\le\ln{(n-1)}\le \ln{n}-\frac{1}{n} .
\]
\begin{proof}
The lemma follows directly from the fact that
\[
\ln{n}-\ln{(n-1)}=\frac{1}{\xi}
\]
for some $\xi\in (n-1,n)$, which can be obtained through Taylor's Theorem.\qed
\end{proof}

\noindent\textbf{Proposition~\ref{prop:integralapproximation}.}
For any natural number $n\geq 2$, we have:
\begin{equation}\label{eq:reciprocalapprox}
 \int_1^n \frac{1}{x}\,\mathrm{d}x-\sum_{j=1}^{n-1} \frac{1}{j}\in \left[-0.7552,-\frac{1}{6}\right]
\end{equation}
\begin{equation}\label{eq:logarithmapprox}
\int_1^n \ln{x}\,\mathrm{d}x-\left(\sum_{j=1}^{n-1} \ln{j}\right) - \frac{1}{2}\cdot \int_1^n \frac{1}{x}\,\mathrm{d}x\in \left[-\frac{1}{12}, 0.2701\right]
\end{equation}
\begin{equation}\label{eq:xlogxapprox}
\int_1^n x\cdot \ln{x}\,\mathrm{d}x-\left(\sum_{j=1}^{n-1} j\cdot\ln{j}\right)-\frac{1}{2}\cdot\int_1^n \ln{x}\,\mathrm{d}x+\frac{1}{12}\cdot \int_1^n \frac{1}{x}\,\mathrm{d}x-\frac{n-1}{2}\in \left[-\frac{19}{72},0.1575\right].
\end{equation}
\begin{proof}
Let $n$ be a natural number such that $n\ge 2$.
We first estimate the difference
\[
\int_1^n \frac{1}{x}\,\mathrm{d}x-\sum_{j=1}^{n-1} \frac{1}{j}~~.
\]
To this end, we deduce the following equalities:
\begin{align*}
 &\int_1^n \frac{1}{x}\,\mathrm{d}x-\sum_{j=1}^{n-1} \frac{1}{j} \\
= &\sum_{j=1}^{n-1}\int_j^{j+1} \left[\frac{1}{x}-\frac{1}{j}\right]\,\mathrm{d}x \\
= &\sum_{j=1}^{n-1}\int_0^{1} \left[\frac{1}{j+x}-\frac{1}{j}\right]\,\mathrm{d}x \\
= & \sum_{j=1}^{n-1}\int_0^{1} \left[-\frac{1}{j^2}\cdot x +\frac{1}{\xi^3_{j,x}}\cdot x^2\right]\,\mathrm{d}x\\
= & -\frac{1}{2}\cdot \left(\sum_{j=1}^{n-1}\frac{1}{j^2}\right)+ \sum_{j=1}^{n-1}\int_0^{1}\frac{1}{\xi^3_{j,x}}\cdot x^2\,\mathrm{d}x~~,\\
\end{align*}
where $\xi_{j,x}$ is a real number in $(j, j+x)$ obtained from Taylor's Theorem with Lagrange's Remainder.
The first and fourth equalities come from the linear property of Riemann Integral;
the second one follows from the variable substitution $x'=x-j$;
the third one follows from Taylor's Theorem.
Using the fact that $\xi_{j,x}\in (j, j+1)$, one obtains that
\begin{equation}\label{eq:reciprocalupperapprox}
\int_1^n \frac{1}{x}\,\mathrm{d}x-\sum_{j=1}^{n-1} \frac{1}{j} \le  -\frac{1}{2}\cdot \left(\sum_{j=1}^{n-1}\frac{1}{j^2}\right) + \frac{1}{3}\cdot \left(\sum_{j=1}^{n-1}\frac{1}{j^3}\right)
\end{equation}
and
\begin{equation}\label{eq:reciprocallowerapprox}
\int_1^n \frac{1}{x}\,\mathrm{d}x-\sum_{j=1}^{n-1} \frac{1}{j} \ge  -\frac{1}{2}\cdot \left(\sum_{j=1}^{n-1}\frac{1}{j^2}\right) + \frac{1}{3}\cdot \left(\sum_{j=2}^{n}\frac{1}{j^3}\right).
\end{equation}
Then (\ref{eq:reciprocalapprox}) follows from the facts that
\begin{align*}
\int_1^n \frac{1}{x}\,\mathrm{d}x-\sum_{j=1}^{n-1} \frac{1}{j} & \le   \sum_{j=1}^{n-1}\left(-\frac{1}{2\cdot j^2} + \frac{1}{3\cdot j^3}\right)\\
& \le  -\frac{1}{2\cdot 1^2} + \frac{1}{3\cdot 1^3}\\
& =-\frac{1}{6}
\end{align*}
and
\begin{align*}
\int_1^n \frac{1}{x}\,\mathrm{d}x-\sum_{j=1}^{n-1} \frac{1}{j} & \ge   \sum_{j=1}^{n-1}\left(-\frac{1}{2\cdot j^2} + \frac{1}{3\cdot j^3}\right)-\frac{1}{3}+\frac{1}{3\cdot n^3}\\
& \ge  -\frac{\pi^2}{12}+\frac{\alpha}{3}-\frac{1}{3}\\
& \ge -0.7552
\end{align*}
where in both situations we use the fact that $2\cdot j^2\le 3\cdot j^3$ for all $j\in\Nset$.

Then we consider the difference
\[
\int_1^n \ln{x}\,\mathrm{d}x-\sum_{j=1}^{n-1} \ln{j}~.
\]
First, we derive that
\begin{align*}
 &\int_1^n \ln{x}\,\mathrm{d}x-\sum_{j=1}^{n-1} \ln{j} \\
= &\sum_{j=1}^{n-1}\int_j^{j+1} \left[\ln{x}-\ln{j}\right]\,\mathrm{d}x \\
= &\sum_{j=1}^{n-1}\int_0^{1} \left[\ln{(j+x)}-\ln{j}\right]\,\mathrm{d}x \\
= & \sum_{j=1}^{n-1}\int_0^{1} \left[\frac{1}{j}\cdot x -\frac{1}{2\cdot \xi^2_{j,x}}\cdot x^2\right]\,\mathrm{d}x\\
= & \frac{1}{2}\cdot \left(\sum_{j=1}^{n-1}\frac{1}{j}\right)- \sum_{j=1}^{n-1}\int_0^{1}\frac{1}{2\cdot \xi^2_{j,x}}\cdot x^2\,\mathrm{d}x \\
\end{align*}
where $\xi_{j,x}$ is a real number in $(j, j+1)$ obtained from Taylor's Theorem.
Using the fact that $\xi_{j,x}\in (j, j+1)$, one can obtain that
\begin{align}\label{eq:logarithmupperapprox}
 & \int_1^n \ln{x}\,\mathrm{d}x-\sum_{j=1}^{n-1} \ln{j} \\
\le & \frac{1}{2}\cdot \left(\sum_{j=1}^{n-1}\frac{1}{j}\right)-\frac{1}{6}\cdot\sum_{j=2}^{n}\frac{1}{j^2}\nonumber \\
\le  & \frac{1}{2}\cdot \int_1^n \frac{1}{x}\,\mathrm{d}x + \frac{1}{4}\cdot \left(\sum_{j=1}^{n-1}\frac{1}{j^2}\right) - \frac{1}{6}\cdot \left(\sum_{j=2}^{n}\frac{1}{j^3}\right)\nonumber\\
&\qquad{}-\frac{1}{6}\cdot\sum_{j=2}^{n}\frac{1}{j^2} \nonumber \\
= & \frac{1}{2}\cdot \int_1^n \frac{1}{x}\,\mathrm{d}x + \sum_{j=1}^{n-1}\left(\frac{1}{12\cdot j^2}-\frac{1}{6\cdot j^3}\right)\nonumber\\
 & \qquad{}+\frac{1}{3}-\frac{1}{6\cdot n^3}-\frac{1}{6\cdot n^2} \nonumber
\end{align}
where the second inequality follows from Inequality~(\ref{eq:reciprocallowerapprox}), and
\begin{align}\label{eq:logarithmlowerapprox}
& \int_1^n \ln{x}\,\mathrm{d}x-\sum_{j=1}^{n-1} \ln{j} \\
\ge & \frac{1}{2}\cdot \left(\sum_{j=1}^{n-1}\frac{1}{j}\right)-\frac{1}{6}\cdot\sum_{j=1}^{n-1}\frac{1}{j^2}\nonumber\\
\ge & \frac{1}{2}\cdot\int_1^n \frac{1}{x}\,\mathrm{d}x+\frac{1}{12}\cdot \left(\sum_{j=1}^{n-1}\frac{1}{j^2}\right) - \frac{1}{6}\cdot \left(\sum_{j=1}^{n-1}\frac{1}{j^3}\right)\nonumber\\
= & \frac{1}{2}\cdot\int_1^n \frac{1}{x}\,\mathrm{d}x + \sum_{j=1}^{n-1}\left(\frac{1}{12\cdot j^2}-\frac{1}{6\cdot j^3}\right)\nonumber
\end{align}
where the second inequality follows from Inequality~(\ref{eq:reciprocalupperapprox}).
Then from Inequality~(\ref{eq:logarithmupperapprox}) and Inequality~(\ref{eq:logarithmlowerapprox}), one has that
\begin{align*}
& \int_1^n \ln{x}\,\mathrm{d}x-\sum_{j=1}^{n-1} \ln{j} \\
\le &  \frac{1}{2}\cdot\int_1^n \frac{1}{x}\,\mathrm{d}x+\sum_{j=1}^{\infty}\left(\frac{1}{12\cdot j^2}-\frac{1}{6\cdot j^3}\right)+\frac{1}{3} \\
\le & \frac{1}{2}\cdot \int_1^n \frac{1}{x}\,\mathrm{d}x+\frac{\pi^2}{72}-\frac{\alpha}{6}+\frac{1}{3} \\
\le & \frac{1}{2}\cdot \int_1^n \frac{1}{x}\,\mathrm{d}x+0.2701
\end{align*}
and
\begin{align*}
& \int_1^n \ln{x}\,\mathrm{d}x-\sum_{j=1}^{n-1} \ln{j} \\
\ge & \frac{1}{2}\cdot\int_1^n \frac{1}{x}\,\mathrm{d}x+\left(\frac{1}{12}-\frac{1}{6}\right)\\
\ge & \frac{1}{2}\cdot\int_1^n \frac{1}{x}\,\mathrm{d}x -\frac{1}{12}
\end{align*}
where in both situations we use the fact that $12\cdot j^2\le 6\cdot j^3$ for all $j\ge 2$. The inequalities above directly imply the inequalities in (\ref{eq:logarithmapprox}).
Finally, we consider the difference
\[
\int_1^n x\cdot \ln{x}\,\mathrm{d}x-\sum_{j=m}^{n-1} j\cdot\ln{j}~~.
\]
Following similar approaches, we derive that for all natural numbers $n\ge 2$,
\begin{align*}
 &\int_1^n x\cdot \ln{x}\,\mathrm{d}x-\sum_{j=1}^{n-1} j\cdot\ln{j} \\
= &\sum_{j=1}^{n-1}\int_j^{j+1} \left[x\cdot \ln{x}-j\cdot\ln{j}\right]\,\mathrm{d}x \\
= &\sum_{j=1}^{n-1}\int_0^{1} \left[(j+x)\cdot \ln{(j+x)}-j\cdot\ln{j}\right]\,\mathrm{d}x \\
= & \sum_{j=1}^{n-1}\int_0^{1} \left[(\ln{j}+1)\cdot x +\frac{1}{2\cdot \xi_{j,x}}\cdot x^2\right]\,\mathrm{d}x\\
= & \frac{1}{2}\cdot \left(\sum_{j=1}^{n-1}\ln{j}\right)+\frac{n-1}{2}+ \sum_{j=1}^{n-1}\int_0^{1}\frac{1}{2\cdot \xi_{j,x}}\cdot x^2\,\mathrm{d}x \\
\end{align*}
where  $\xi_{j,x}\in (j, j+1)$. Thus, one obtains that
\begin{align}\label{eq:xlogxupperapprox}
& \int_1^n x\cdot \ln{x}\,\mathrm{d}x-\sum_{j=1}^{n-1} j\cdot\ln{j} \le \\
& \qquad\frac{1}{2}\cdot \left(\sum_{j=1}^{n-1}\ln{j}\right)+\frac{n-1}{2}+ \frac{1}{6}\cdot\sum_{j=1}^{n-1}\frac{1}{j}\nonumber
\end{align}
and
\begin{align}\label{eq:xlogxlowerapprox}
&\int_1^n x\cdot \ln{x}\,\mathrm{d}x-\sum_{j=1}^{n-1} j\cdot\ln{j}\ge \\
&\qquad\frac{1}{2}\cdot \left(\sum_{j=1}^{n-1}\ln{j}\right)+\frac{n-1}{2}+\frac{1}{6}\cdot\sum_{j=1}^{n-1}\frac{1}{j}-\frac{1}{6}+\frac{1}{6\cdot n}.\nonumber
\end{align}
By plugging Inequalities in~(\ref{eq:reciprocallowerapprox}) and~(\ref{eq:logarithmlowerapprox}) into Inequality~(\ref{eq:xlogxupperapprox}), one obtains that
\begin{align*}
& \int_1^n x\cdot \ln{x}\,\mathrm{d}x-\sum_{j=1}^{n-1} j\cdot\ln{j} \\
\le & \frac{1}{2}\cdot\left[\int_1^n \ln{x}\,\mathrm{d}x-\frac{1}{2}\cdot\int_1^n \frac{1}{x}\,\mathrm{d}x - \sum_{j=1}^{n-1}\left(\frac{1}{12\cdot j^2}-\frac{1}{6\cdot j^3}\right)\right] \\
 & {}+\frac{n-1}{2}\\
 & {}+\frac{1}{6}\cdot\left[\int_1^n \frac{1}{x}\,\mathrm{d}x +\frac{1}{2}\cdot \left(\sum_{j=1}^{n-1}\frac{1}{j^2}\right) - \frac{1}{3}\cdot \left(\sum_{j=2}^{n}\frac{1}{j^3}\right)\right] \\
 \le & \frac{1}{2}\cdot\int_1^n \ln{x}\,\mathrm{d}x-\frac{1}{12}\cdot \int_1^n \frac{1}{x}\,\mathrm{d}x+\frac{n-1}{2}\\
  & {}+\sum_{j=1}^{n-1}\left(\frac{1}{24\cdot j^2}+\frac{1}{36\cdot j^3}\right)+\frac{1}{18}-\frac{1}{18\cdot n^3}  \\
\le & \frac{1}{2}\cdot\int_1^n \ln{x}\,\mathrm{d}x-\frac{1}{12}\cdot \int_1^n \frac{1}{x}\,\mathrm{d}x+\frac{n-1}{2}+\frac{\pi^2}{144}+\frac{\alpha}{36}+\frac{1}{18} \\
\le & \frac{1}{2}\cdot\int_1^n \ln{x}\,\mathrm{d}x-\frac{1}{12}\cdot \int_1^n \frac{1}{x}\,\mathrm{d}x+\frac{n-1}{2} + 0.1575
\end{align*}
for all natural numbers $n\ge 2$.
Similarly, by plugging Inequalities in (\ref{eq:reciprocalupperapprox}) and (\ref{eq:logarithmupperapprox}) into Inequality~(\ref{eq:xlogxlowerapprox}), one obtains
\begin{align*}
& \int_1^n x\cdot \ln{x}\,\mathrm{d}x-\sum_{j=1}^{n-1} j\cdot\ln{j} \\
\ge & \frac{1}{2}\cdot\left[\int_1^n \ln{x}\,\mathrm{d}x-\frac{1}{2}\cdot\int_1^n \frac{1}{x}\,\mathrm{d}x - \sum_{j=1}^{n-1}\left(\frac{1}{12\cdot j^2}-\frac{1}{6\cdot j^3}\right)-\frac{1}{3}\right] \\
  & {}+\frac{n-1}{2}\\
  & {}+\frac{1}{6}\cdot\left[\int_1^n \frac{1}{x}\,\mathrm{d}x +\frac{1}{2}\cdot \left(\sum_{j=1}^{n-1}\frac{1}{j^2}\right) - \frac{1}{3}\cdot \left(\sum_{j=1}^{n-1}\frac{1}{j^3}\right)\right]-\frac{1}{6} \\
 \ge & \frac{1}{2}\cdot\int_1^n \ln{x}\,\mathrm{d}x-\frac{1}{12}\cdot \int_1^n \frac{1}{x}\,\mathrm{d}x+\frac{n-1}{2}\\
  & {}+\sum_{j=1}^{n-1}\left(\frac{1}{24\cdot j^2}+\frac{1}{36\cdot j^3}\right)-\frac{1}{3}  \\
\ge & \frac{1}{2}\cdot\int_1^n \ln{x}\,\mathrm{d}x-\frac{1}{12}\cdot \int_1^n \frac{1}{x}\,\mathrm{d}x+\frac{n-1}{2}+\frac{1}{24}+\frac{1}{36}-\frac{1}{3}\\
= & \frac{1}{2}\cdot\int_1^n \ln{x}\,\mathrm{d}x-\frac{1}{12}\cdot \int_1^n \frac{1}{x}\,\mathrm{d}x+\frac{n-1}{2} -\frac{19}{72}
\end{align*}
Then the inequalities in (\ref{eq:xlogxapprox}) are clarified.\qed
\end{proof}

\noindent{\em Example~\ref{ex:overapprox}.}
Consider the summation
\[
\displaystyle\sum_{j=\left\lceil\frac{n}{2}\right\rceil}^{n-1}\ln{j}+ \displaystyle\sum_{j=\left\lfloor\frac{n}{2}\right\rfloor}^{n-1} \ln{j}\quad (n\ge 4).
\]
By Proposition~\ref{prop:integralapproximation}, we can over-approximate it as
\[
2\cdot\left(\Gamma_{\ln{\mathfrak{n}}}\left(n\right)+\frac{1}{12}\right)
-\left(\Gamma_{\ln{\mathfrak{n}}}\left(\left\lceil\frac{n}{2}\right\rceil\right)+\Gamma_{\ln{\mathfrak{n}}}\left(\left\lfloor\frac{n}{2}\right\rfloor\right)-0.5402\right)
\]
which is equal to
\begin{align*}
& 2\cdot n\cdot\ln{n}-2\cdot n-\ln{n}-\left\lceil\frac{n}{2}\right\rceil\cdot\ln{\left\lceil\frac{n}{2}\right\rceil}-\left\lfloor\frac{n}{2}\right\rfloor\cdot\ln{\left\lfloor\frac{n}{2}\right\rfloor}\\
&{}+\left\lceil\frac{n}{2}\right\rceil+\left\lfloor\frac{n}{2}\right\rfloor+\frac{\ln{\left\lfloor\frac{n}{2}\right\rfloor}}{2}+\frac{\ln{\left\lceil\frac{n}{2}\right\rceil}}{2}+\frac{1}{6}+0.5402.
\end{align*}
Then using Proposition~\ref{prop:lnflooroverapprox}, we can further obtain the following over-approximation
\begin{align*}
& 2\cdot n\cdot\ln{n}-2\cdot n-\ln{n}+0.7069-\frac{n}{2}\cdot\left(\ln{n}-\ln{2}\right)-\frac{n-1}{2}\cdot\left(\ln{n}-\ln{2}-\frac{1}{n-1}\right)\\
&{}+\frac{n+1}{2}+\frac{n}{2}+\frac{\ln{n}-\ln{2}}{2}+\frac{\ln{n}-\ln{2}+\frac{1}{n}}{2}
\end{align*}
which is roughly
$n\cdot\ln{n}-(1-\ln{2})\cdot n+\frac{1}{2}\cdot\ln{n}+0.6672+\frac{1}{2\cdot n}$.\qed

\section{Proofs for Sect.~\ref{sect:unisynth}}\label{app:unisynth}

\noindent\textbf{Lemma~\ref{lemm:unitrans}.}
Let $\mathfrak{f}\in\{\ln{\mathfrak{n}},\mathfrak{n},\mathfrak{n}\cdot\ln{\mathfrak{n}}\}$
and $c$ be a constant.
For all univariate recurrence expressions $\mathfrak{g}$, there exists pseudo-polynomials
$p$ and $q$ such that coefficients (i.e., $a_i,b_i$'s in~(\ref{eq:pseudopoly}))
of $q$ are all non-negative, $C_q>0$ and the following assertion holds:
for all $d>0$ and for all $n\ge 2$, with $h=d\cdot \Sub({\mathfrak{f}})+c$,
the inequality $\OvAp(\mathfrak{g}, h)(n)\le h(n)$ is equivalent to
$d\cdot p(n)\ge q(n)$.
\begin{proof}
From Definition~\ref{def:unioverapprox}, $n\mapsto n\cdot(n-1)\cdot\OvAp(\mathfrak{g}, h)(n)$ is a pseudo-polynomial.
Simple rearrangement of terms in inequality $\OvAp(\mathfrak{g}, h)(n)\le h(n)$ gives the desired
pseudo-polynomials.
Moreover, the fact that all coefficients in $\mathfrak{g}$ (from~(\ref{eq:unirecurrel})) are positive,
is used to derive that all coefficients of $q$ are non-negative and $C_q>0$.\qed
\end{proof}

\noindent\textbf{Proposition~\ref{prop:unisufflarge}.}
Let $p,q$ be pseudo-polynomials such that $C_q>0$ and
all coefficients of $q$ are non-negative.
Then there exists a real number $d>0$ such that
$d\cdot p(n)\ge q(n)$
for sufficiently large $n$ iff $\mathrm{deg}(p)\ge \mathrm{deg}(q)$ and $C_p>0$.
\begin{proof} We present the two directions of the proof.

(``\emph{If}'':) Suppose that $\mathrm{deg}(p)\ge \mathrm{deg}(q)$ and $C_p>0$.
Then the result follows directly from the facts that
(i) $\frac{q(n)}{p(n)}>0$ for sufficiently large $n$ and (ii)  $\lim\limits_{n\rightarrow\infty}\frac{q(n)}{p(n)}$ exists and is non-negative.

(``\emph{Only-if}'':) Let $d$ be a positive real number such that $d\cdot p(n)\ge q(n)$ for sufficiently large $n$. Then $C_p>0$, or otherwise $d\cdot p(n)$ is either constantly zero or negative for sufficiently large $n$.
Moreover, $\mathrm{deg}(p)\ge \mathrm{deg}(q)$, since otherwise $\lim\limits_{n\rightarrow\infty}\frac{q(n)}{p(n)}=\infty$.\qed
\end{proof}

\noindent\textbf{Proposition~\ref{prop:unisufflargeN}.}
Consider two univariate pseudo-polynomials $p,q$ such that $\mathrm{deg}(p)\ge \mathrm{deg}(q)$, all coefficients of $q$ are non-negative and $C_p,C_q>0$.
Then given any $\epsilon\in (0,1)$,
\[
\frac{q(n)}{p(n)}\le \frac{\mathbf{1}_{\mathrm{deg}(p)=\mathrm{deg}(q)}\cdot \frac{C_q}{C_p}+\epsilon}{1-\epsilon}
\]
for all $n\ge N_{\epsilon,p,q}$ (for $N_{\epsilon,p,q}$ of Definition~\ref{def:unisuffN}).
\begin{proof}
Let $p,q$ be given in Definition~\ref{def:unisuffN}.
Fix an arbitrary $\epsilon\in (0,1)$ and let $N_{\epsilon,p,q}$ be given in Definition~\ref{def:unisuffN}.
Then for all $n\ge N_{\epsilon,p,q}$, (i) both $p(n),q(n)$ are positive and (ii)
\begin{align*}
\frac{q(n)}{\overline{p}(n)} &\le \sum_{i=1}^{k'} a'_i\cdot \frac{N^{i}\cdot \ln{N}}{\overline{p}(N)}+\sum_{i=1}^{\ell'} b'_i\cdot \frac{N^{i}}{\overline{p}(N)}\\
&\le \mathbf{1}_{\mathrm{deg}(p)=\mathrm{deg}(q)}\cdot\frac{C_q}{C_p}+\epsilon
\end{align*}
and
\begin{align*}
\frac{p(n)}{\overline{p}(n)} &\ge 1-\left[-1+\sum_{i=1}^{k} |a_i|\cdot \frac{N^{i}\cdot \ln{N}}{\overline{p}(N)}+\sum_{i=1}^{\ell} |b_i|\cdot \frac{N^{i}}{\overline{p}(N)}\right]\\
&\ge 1-\epsilon.
\end{align*}
It follows that for all $n\ge N_{\epsilon,p,q}$,
\[
\frac{q(n)}{p(n)}\le \frac{\mathbf{1}_{\mathrm{deg}(p)=\mathrm{deg}(q)}\cdot \frac{C_q}{C_p}+\epsilon}{1-\epsilon}~~.
\]
The desired result follows.\qed
\end{proof}

\noindent\textbf{Theorem~\ref{thm:soundnessunidec}.}[Soundness for $\mbox{\sl UniDec}$]
If $\mbox{\sl UniDec}$ outputs ``$\mbox{\sl yes}$'', then there exists a univariate guess-and-check function in form~(\ref{eq:uniguess})
for the inputs $G$ and $\mathfrak{f}$.
The algorithm is a linear-time algorithm in the size of the input recurrence relation.
\begin{proof}
From Definition~\ref{def:uniguess} and the special form (\ref{eq:uniguess}) for univariate guess-and-check functions, a function in form (\ref{eq:uniguess}) which satisfies the inductive argument of Definition~\ref{def:uniguess} can be modified to satisfy also the base condition of Definition~\ref{def:uniguess} by simply raising $d$ to a sufficiently large amount.
Then the correctness of the algorithm follows from Theorem~\ref{thm:uniguess} and the sufficiency of Proposition~\ref{prop:unisufflarge}.
Furthermore, the algorithm runs in linear time since the transformation from the inequality $\OvAp(\mathfrak{g}, h)(n)\le h(n)$ into $d\cdot p(n)\ge q(n)$ (cf. Lemma~\ref{lemm:unitrans}) takes linear time in the size of the input recurrence relation.
\qed
\end{proof}

\smallskip
\noindent\textbf{Theorem~\ref{thm:soundnessunisynth}.}[Soundness for $\mbox{\sl UniSynth}$]
If the algorithm $\mbox{\sl UniSynth}$ outputs a real number $d$, then $d\cdot \Sub(\mathfrak{f})+c$ is a univariate guess-and-check function for $G$.
\begin{proof}
Directly from the construction of the algorithm, Theorem~\ref{thm:uniguess}, Proposition~\ref{prop:unisufflarge} and Proposition~\ref{prop:unisufflargeN}.\qed
\end{proof}

\section{Detailed Experimental Results}\label{app:experiments}

The detailed experimental results are given in Table~\ref{tbl:detailedexperiments}.
We use $\checkmark$ to represent $\mbox{\sl yes}$ and $\times$ for $\mbox{\sl fail}$.
In addition to Table~\ref{tab:experiments}, we include values for $N_{\epsilon,p,q}$ in Definition~\ref{def:unisuffN}.
For the separable bivariate examples, recall that $n$ does not change, and in these examples, the reduction
to the univariate case is the function of $m$.

\begin{table*}
\centering
\begin{tabular}{ |c|c|c|c|c|c|c| }
\hline
\multirow{2}{*}{\sc Program} & \multirow{2}{*}{\sc $\mathfrak{f}$} & \multirow{2}{*}{\sc UniDec  } & \multicolumn{4}{|c|}{\sc UniSynth(\checkmark)}\\
\cline{4-7}
& & & $\epsilon$ & $N_{\epsilon,p,q}$ & $d$ & $d_{100}$ \\
\hline
\hline
\multirow{4}{*}{\sc R.-Sear.} & \multirow{4}{*}{\sc $\ln \mathfrak{n}$}
& \multirow{4}{*}{\sc \checkmark} & $0.5$ & $13$ & $40.107$ & \multirow{4}{*}{\sc $15.129$ } \\
\cline{4-6}
&&& $0.3$ & $25$ & $28.363$ &  \\
\cline{4-6}
&&& $0.1$ & $97$  & $21.838$ & \\
\cline{4-6}
&&& $0.01$ & $1398$ & $19.762$ &\\
\hline
\hline
\multirow{6}{*}{\sc Q.-Sort} & $\ln \mathfrak{n}$ & $\times$ & \multirow{2}{*}{\sc -} & \multirow{2}{*}{\sc -} & \multirow{2}{*}{\sc -} &\multirow{2}{*}{\sc -} \\
\cline{2-3}
& $\mathfrak{n}$ & $\times$ & & & &   \\
\cline{2-7}
& \multirow{4}{*}{\sc $\mathfrak{n}\ln \mathfrak{n}$}
& \multirow{4}{*}{\sc \checkmark}  & $0.5$ & $10$ &  $9.001$ & \multirow{4}{*}{\sc $3.172$ } \\
\cline{4-6}
&&& $0.3$ & $21$    &  $6.143$ &  \\
\cline{4-6}
&&& $0.1$ & $91$  &  $4.556$ & \\
\cline{4-6}
&&& $0.01$ & $1458$  & $4.051$ &\\
\hline
\hline
\multirow{5}{*}{\sc Q.-Select} & $\ln \mathfrak{n}$ & $\times$ & - & - & - & - \\
\cline{2-7}
& \multirow{4}{*}{\sc $\mathfrak{n}$}
& \multirow{4}{*}{\sc \checkmark}  & $0.5$ & $33$ &  $17.001$ & \multirow{4}{*}{\sc $7.909$ } \\
\cline{4-6}
&&& $0.3$ & $54$  & $11.851$ &  \\
\cline{4-6}
&&& $0.1$ & $160$ &  $9.001$ & \\
\cline{4-6}
&&& $0.01$ & $1600$  &  $8.091$ &\\
\hline
\hline
\multirow{6}{*}{\sc Diam. A} & $\ln \mathfrak{n}$ & $\times$ & \multirow{2}{*}{\sc -} & \multirow{2}{*}{\sc -} & \multirow{2}{*}{\sc -} & \multirow{2}{*}{\sc -}  \\
\cline{2-3}
& $\mathfrak{n}$ & $\times$ & & & &   \\
\cline{2-7}
& \multirow{4}{*}{\sc $\mathfrak{n}\ln \mathfrak{n}$}
& \multirow{4}{*}{\sc \checkmark} &  $0.5$ & $3$ &  $9.001$ & \multirow{4}{*}{\sc $4.525$ } \\
\cline{4-6}
&&& $0.3$ & $3$        &       $6.143$ &  \\
\cline{4-6}
&&& $0.1$ & $4$ &      $4.556$ & \\
\cline{4-6}
&&& $0.01$ & $4$ &     $4.525$ &\\
\hline
\hline
\multirow{5}{*}{\sc Diam. B} & $\ln \mathfrak{n}$ & $\times$ & - & - & - & - \\
\cline{2-7}
& \multirow{4}{*}{\sc $\mathfrak{n}$}
& \multirow{4}{*}{\sc \checkmark}  & $0.5$ & $9$  & $13.001$ & \multirow{4}{*}{\sc $5.918$ } \\
\cline{4-6}
&&& $0.3$ & $14$ &     $9.001$ &  \\
\cline{4-6}
&&& $0.1$ & $40$ &     $6.778$ & \\
\cline{4-6}
&&& $0.01$ & $400$ & $6.071$ &\\
\hline
\hline
\multirow{6}{*}{\sc Sort-Sel.} & $\ln \mathfrak{n}$ & $\times$ & \multirow{2}{*}{\sc -} & \multirow{2}{*}{\sc -} & \multirow{2}{*}{\sc -} & \multirow{2}{*}{\sc -}  \\
\cline{2-3}
& $\mathfrak{n}$ & $\times$ & & & &   \\
\cline{2-7}
& \multirow{4}{*}{\sc $\mathfrak{n}\ln \mathfrak{n}$}
& \multirow{4}{*}{\sc \checkmark} &  $0.5$ & $18$  & $50.052$ & \multirow{4}{*}{\sc $16.000$} \\
\cline{4-6}
&&& $0.3$ & $29$ &     $24.852$ &  \\
\cline{4-6}
&&& $0.1$ & $87$ & $17.313$ & \\
\cline{4-6}
&&& $0.01$ & $866$ &   $16.000$ & \\
\hline
\hline
\multirow{4}{*}{\sc Coupon} & \multirow{4}{*}{\sc $\mathfrak{n}\cdot\ln \mathfrak{m}$}
& \multirow{4}{*}{\sc \checkmark}  & $0.5$ & $2$  &  $3.001$ & \multirow{4}{*}{\sc $0.910$ } \\
\cline{4-6}
&&& $0.3$ & $2$ &      $1.858$ &  \\
\cline{4-6}
&&& $0.1$ & $2$ &      $1.223$ & \\
\cline{4-6}
&&& $0.01$ & $2$ &     $1.021$ &\\
\hline
\hline
\multirow{4}{*}{\sc Res. A} & \multirow{4}{*}{\sc $\mathfrak{n}\cdot\ln \mathfrak{m}$}
& \multirow{4}{*}{\sc \checkmark}  & $0.5$ & $2$     &       $6.437$ & \multirow{4}{*}{\sc $2.472$ } \\
\cline{4-6}
&&& $0.3$ & $2$ &      $4.312$ &  \\
\cline{4-6}
&&& $0.1$ & $2$ &      $3.132$ & \\
\cline{4-6}
&&& $0.01$ & $2$ &     $2.756$ &\\
\hline
\hline
\multirow{5}{*}{\sc Res. B} & $\ln \mathfrak{m}$ & $\times$ & - & - & - & -\\
\cline{2-7}
&\multirow{4}{*}{\sc $\mathfrak{m}$}
& \multirow{4}{*}{\sc \checkmark} & $0.5$ & $2$ &  $6.437$ & \multirow{4}{*}{\sc $2.691$ } \\
\cline{4-6}
&&& $0.3$ & $2$ &      $4.312$ &  \\
\cline{4-6}
&&& $0.1$ & $2$ &      $3.132$ & \\
\cline{4-6}
&&& $0.01$ & $2$ &     $2.756$ &\\
\hline
\end{tabular}
\caption{Detailed experimental results where all running times (averaged over $5$ runs) are less than $0.02$ seconds
(between $0.01$ and $0.02$ seconds).}
\label{tbl:detailedexperiments}
\end{table*}